\documentclass[twocolumn,showpacs,preprintnumbers]{revtex4}
\usepackage{amssymb}
\usepackage{amsfonts}
\usepackage{amsmath}
\usepackage{graphicx}
\usepackage{dcolumn}
\usepackage{bm}
\usepackage{epsfig}

\begin{document}

\preprint{APS/123-QED}
\title{Theory of spatially non-symmetric kinetic equilibria for
collisionless plasmas}
\author{Claudio Cremaschini}
\affiliation{Department of Mathematics and Geosciences, University
of Trieste, Via Valerio 12, 34127 Trieste, Italy}
\affiliation{Institute of Physics, Faculty of Philosophy and
Science, Silesian University in Opava, Bezru\v{c}ovo n\'{a}m.13,
CZ-74601 Opava, Czech Republic}
\author{Massimo Tessarotto}
\affiliation{Department of Mathematics and Geosciences, University
of Trieste, Via Valerio 12, 34127 Trieste, Italy}
\date{\today }

\begin{abstract}
The problem posed by the possible existence/non-existence of spatially
non-symmetric kinetic equilibria has remained unsolved in plasma theory. For
collisionless magnetized plasmas this involves the construction of
stationary solutions of the Vlasov-Maxwell equations. In this paper the
issue is addressed for non-relativistic plasmas both in astrophysical and
laboratory contexts. The treatment is based on a Lagrangian variational
description of single-particle dynamics. Starting point is a
non-perturbative formulation of gyrokinetic theory, which allows one to
construct \textquotedblleft a posteriori\textquotedblright\ with prescribed
order of accuracy an asymptotic representation for the magnetic moment. In
terms of the relevant particle adiabatic invariants generalized
bi-Maxwellian equilibria are proved to exist. These are shown to recover,
under suitable assumptions, a Chapman-Enskog form which permits an
analytical treatment of the corresponding fluid moments. In particular, the
constrained posed by the Poisson and the Ampere equations are analyzed, both
for quasi-neutral and non-neutral plasmas. The conditions of existence of
the corresponding non-symmetric kinetic equilibria are investigated. As a
notable feature, both astrophysical and laboratory plasmas are shown to
exhibit, under suitable conditions, a kinetic dynamo, whereby the
equilibrium magnetic field can be self-generated by the\ equilibrium plasma
currents.
\end{abstract}

\maketitle

\bigskip










\bigskip

\section{Introduction}

A fundamental issue of plasma kinetic theory is related to the search of
possible spatially non-symmetric (in particular, non-axisymmetric)
Vlasov-Maxwell kinetic equilibria or more generally quasi-stationary
solutions characterizing magnetized plasmas. As it is well known, the
problem has remained unsolved since its foundation laid down originally by
Harold Grad in a series of famous papers \cite{Grad84,Grad67,Grad58}. In
fact, according to what was pointed out by H. Grad himself, it is very
unlikely that there exists a general class of toroidal non-symmetric
equilibria with nested magnetic surfaces and smooth pressure profiles. His
conclusion was based on the existence of periodicity constraints on the
equilibrium magnetic field which he found to be generally not satisfied. The
constraints considered by Grad had actually been pointed out\ independently
by Kruskal, Kulsrud and Newcomb\ \cite{kuls,kuls2,Newcomb}. These were shown
by Tessarotto \textit{et al.} \cite{Tex95,tex96} to be necessary conditions
for the existence of quasi-symmetric MHD equilibria, i.e., localized
stationary solutions of the scalar-pressure ideal-MHD equations in which the
magnetic field satisfies locally the periodicity constraints on a suitable
set of rational magnetic surfaces. In these works only the case of fluid
equilibria was considered. However, the general problem of the construction
of consistent kinetic (i.e., Vlasov-Maxwell) equilibria was actually not
addressed. Surprisingly, despite its fundamental conceptual relevance for
theoretical plasma physics, the problem has remained unsolved to date.

A further motivation is provided by recent developments regarding the
construction of kinetic equilibria for collisionless axisymmetric magnetized
plasmas \cite{Cr2010,Cr2011,Cr2011a,Cr2012} and their stability properties
\cite{PRL}. In these works it was shown that consistent solutions of the
Vlasov-Maxwell equations can be determined for the kinetic distribution
function (KDF), based on the identification of the relevant single-particle
invariants. The perturbative kinetic theory developed therein permits to
represent the species KDF in terms of generalized Maxwellian distributions,
characterized by temperature anisotropy, non-uniform fluid fields and local
plasma flows. Basic feature of these equilibria is that they can also
support a kinetic dynamo mechanism, which can be responsible for the
self-generation of the electromagnetic (EM) field in which the plasma is
immersed \cite{Cr2010,Cr2011,latvia}. These conclusions apply to
axisymmetric magnetized plasmas both in the laboratory and astrophysical
contexts. Important examples for the two cases are provided by Tokamak
devices and accretion discs respectively.

The question arises, however, whether similar physical mechanisms can be
found for collisionless magnetized plasmas also in the absence of spatial
symmetries. A consistent theoretical description of these systems which may
explain the observed phenomenology should be sought within the framework of
a kinetic theory. This refers in particular, to the possible kinetic origin
of equilibrium flows and electric current densities, together with the
simultaneous occurrence of temperature and pressure anisotropies. The goal
of the present investigation lies in the extension of the kinetic treatment
presented in Refs.\cite{Cr2010,Cr2011,Cr2011a,Cr2012} to generally spatially
non-symmetric plasma configurations in both laboratory and astrophysical
contexts. More precisely, here we refer specifically to the case of local
configurations in bounded domains which are characterized by
quasi-stationary plasma flows (flow-through plasma equilibria). In all cases
plasmas are considered in spatially non-symmetric smooth electromagnetic
fields, with astrophysical plasmas being possibly subject also to the action
of a gravitational field. For this purpose, a necessary prerequisite
consists in the determination of the single-particle invariants occurring in
magnetized plasmas. This goal is reached in the framework of the gyrokinetic
(GK) treatment of single-particle dynamics formulated in the framework of a
variational Lagrangian approach. As a result, the GK magnetic moment is
determined correct to first-order in the Larmor-radius expansion (LRE, see
definition below). The expression of the magnetic moment thus obtained is
then used for the construction of generalized Maxwellian solutions of the
Vlasov equation. A remarkable feature of these equilibria lies in the fact
that they can sustain non-uniform number density and anisotropic
temperatures, while at the same time allowing for parallel and perpendicular
flows and current densities. The latter could be in principle responsible
for the self-generation of stationary non-symmetric magnetic and electric
fields, thus giving rise to corresponding non-symmetric MHD equilibria
(kinetic-MHD equilibria).

The plan of the paper is as follows. In Section 2 the solution method
adopted here for the construction of kinetic equilibria is first discussed.
In Section 3 fundamentals of GK theory are recalled in the framework of the
Hamilton variational principle. In Section 4 a variational non-perturbative
GK theory is developed for energy-conserving systems. After constructing a
GK Lagrangian differential form, conditions of existence for the GK
transformation are obtained together with a non-perturbative expression for
the particle magnetic moment. On this basis, in Section 5 an asymptotic
formulation of GK theory is derived, based on the Larmor-radius expansion.
This allows one to obtain a representation of the magnetic moment in terms
of an adiabatic invariant of first-order in the expansion parameter. Section
6 deals with the construction of exact equilibrium solution of the Vlasov
equation, which are proved to admit a representation in terms of generalized
bi-Maxwellian distributions. A suitable perturbative theory is developed in
Section 7, which permits the analytical treatment of these solutions and
their asymptotic representation in terms of a Chapman-Enskog series
performed around an equilibrium bi-Maxwellian distribution. The equilibrium
fluid moments corresponding to this solution are then calculated
analytically in Section 8, which are identified with the species number
density, flow velocity, pressure tensor and current density. Analysis of the
constraints posed on the equilibrium KDF by the Poisson and the Ampere
equations are investigated respectively in Sections 9 and 10 for both
astrophysical and laboratory plasmas. Concluding remarks are presented in
Section 11, while the Appendix reports details of the Hamilton variational
principles and the corresponding Euler-Lagrange equations in GK variables.

\bigskip

\section{Solution method}

The method adopted for the construction of Vlasov-Maxwell equilibria follows
the guidelines pointed out in a series of recent papers \cite%
{Cr2010,Cr2011,Cr2011a}. The basic requirement is the search of equilibrium
KDFs $f_{s}$ for collisionless plasmas such that, in a suitable subset of
phase-space, each of them can be realized in terms of appropriate
generalized Gaussian distributions able to sustain kinetic-MHD equilibria.
In the case of stationary or quasi-stationary configurations, besides the
Chapman-Enskog method (see for example Refs.\cite{Ma2,Ma3,Ma4,Ma1,Ma5} for
symmetric systems), a convenient solution technique for the Vlasov equation
consists in the determination of particular solutions of the form $%
f_{s}=f_{\ast s}$, where $f_{\ast s}$\ is considered as a function of a
suitable\ set of independent particle invariant phase-functions $\left\{
K_{j},j=1,n\right\} $. More precisely, let $\mathbf{x}\equiv \left( \mathbf{r%
},\mathbf{v}\right) $ be the particle Newtonian state spanning the
phase-space $\Upsilon $ and $\mathbf{z}=\mathbf{z}\left( \mathbf{x},t\right)
$ a suitable transformed state in $\Upsilon ^{\prime }$ defined by the
phase-space diffeomorphism $\Upsilon \rightarrow \Upsilon ^{\prime }:\mathbf{%
x}\leftrightarrow \mathbf{z}\left( \mathbf{x},t\right) $. Then, the local
phase-functions $K_{j}=K_{j}\left( \mathbf{z},t\right) $ are to be
identified with exact or adiabatic invariants, here referred to as pointwise
invariants, which satisfy one of the following conditions:

a) \textit{Local (global) first integral: }$K_{j}$ does not depend
explicitly on time and moreover the equation%
\begin{equation}
\frac{d}{dt}K_{j}\left( \mathbf{z}\right) =0  \label{fi}
\end{equation}%
holds for all $\mathbf{z}$ belonging to the proper subset of $\Upsilon
_{1}^{\prime }\subset \Upsilon ^{\prime }$ (respectively, $\Upsilon
_{1}^{\prime }=\Upsilon ^{\prime }$).

b) \textit{Local (global) adiabatic invariant of order }$k\geq 1$\textit{: }$%
K_{j}$ can depend at most slowly on time, in the sense $K_{j}=K_{j}\left(
\mathbf{z},\varepsilon ^{k}t\right) $, and satisfies the asymptotic equation%
\begin{equation}
\frac{d}{dt}K_{j}\left( \mathbf{z},\varepsilon ^{k}t\right) =0+O\left(
\varepsilon ^{k}\right) ,  \label{ai}
\end{equation}%
for all $\mathbf{z}$ belonging to the proper subset of $\Upsilon
_{1}^{\prime }\subset \Upsilon ^{\prime }$ (respectively, $\Upsilon
_{1}^{\prime }=\Upsilon ^{\prime }$). Here $\varepsilon \ll 1$ is a suitable
dimensionless parameter to be considered as infinitesimal and later
identified with the Larmor-radius parameter (see Section 5).

If $f_{\ast s}$ is a function of first integrals only, then it is itself a
first integral. This identifies a stationary KDF. On the other hand, if $%
f_{\ast s}$ depends also on adiabatic invariants of prescribed order, it is
necessarily quasi-stationary, in the sense that it is itself an adiabatic
invariant too. As a short way, in the following, in both cases $f_{\ast s}$
will be referred to as an equilibrium KDF (or kinetic equilibrium).

Notice that the notion of adiabatic invariant given by Eq.(\ref{ai}) and
required for the solution method adopted here, generally differs from that
considered previously in some literature. In particular, both Eqs.(\ref{fi})
and (\ref{ai}) state that each $K_{j}$ must remain constant pointwise along
a phase-space trajectory, and not just merely in some averaged sense. In the
latter case the notion of invariant, based on Eq.(\ref{ai}), is replaced by
the requirement that, for a prescribed $k\geq 1$ and $0<\alpha <k$, the
averaged equation $\left\langle \frac{d}{dt}K_{j}\right\rangle =0+O\left(
\varepsilon ^{\alpha }\right) $ holds, with the brackets \textquotedblleft $%
\left\langle {}\right\rangle $\textquotedblright\ denoting a suitable
phase-space average operator (cf gyrophase average; see discussion below in
Section 3). The type of requirement indicated by Eqs.(\ref{fi}) and (\ref{ai}%
) necessarily demands the adoption of gyrokinetic theory. Approaches of this
type are well-known in the literature (see for example the Hamiltonian
Lie-transform method by Dubin et al. \cite{dubin83}, also adopted by Brizard
\cite{brizard95} in the case of laboratory plasmas). In this paper however
we formulate the problem from a more general viewpoint applicable also to
astrophysical plasmas immersed in equilibrium configurations. This is based
on a Lagrangian formulation valid in principle also for finite
guiding-center trasformations. Such a formulation allows one to determine
\textquotedblleft a posteriori\textquotedblright\ explicit perturbative
corrections of the magnetic moment with prescribed accuracy.

To conclude this section, it is important to remark that a basic requirement
for the application of the method is that it should also afford determining
\textquotedblleft a posteriori\textquotedblright\ a perturbative
representation of the KDF equivalent to the Chapman-Enskog expansion,
according to the perturbative technique developed in Refs.\cite%
{Cr2010,Cr2011,Cr2011a,Cr2012} for axisymmetric plasmas. Here we point out
that in fact this goal can be reached also for non-symmetric configurations,
in the framework of validity of GK theory and in a suitable subset of
phase-space characteristic for each plasma species.

\bigskip

\section{Fundamentals of GK theory}

The theory of charged particle dynamics in intense EM fields has a long
history, whose origin dates back to Alfven \cite{Alfven40} who first showed
that the magnetic moment is an adiabatic invariant associated to the fast
rotation of charged particles around magnetic flux lines. The problem
belongs to the more general framework related to the search of local/global
first integrals of motion and/or adiabatic invariants for classical
dynamical systems represented by Newtonian $N$-body systems of charged
particles.

In the context of classical dynamics, both for non-relativistic and
relativistic charged particles, the appropriate theoretical framework is
provided by the Hamilton variational principle \cite{EPJ1,EPJ2,EPJ3,EPJ4}.
In fact, the result follows in a natural way thanks to the validity of
Noether theorem as a consequence of the existence of exact or approximate
symmetry transformations. When EM radiation-reaction effects are neglected,
the treatment in the two cases exhibits analogous qualitative properties. In
particular, it can be shown that, under suitable conditions, the magnetic
moment in both cases remains an adiabatic invariant to arbitrary order in
terms of the asymptotic expansion parameter $\varepsilon $, to be associated
with the particle Larmor radius (see Refs.\cite{kruskal,Tex99} and the
discussion below).

In this paper we focus on the dynamics of isolated point-like charges in
suitably-smooth spatially non-symmetric and stationary EM fields, ignoring
relativistic effects. Thus, in the present context, the phase-space
transformation of interest $\Upsilon \rightarrow \Upsilon ^{\prime }$ is
identified with the so-called gyrokinetic (GK) transformation%
\begin{equation}
\mathbf{x}\equiv \left( \mathbf{r},\mathbf{v}\right) \mathbf{\rightarrow z}%
^{\prime }\equiv \left( \mathbf{y}^{\prime },\phi ^{\prime }\right) ,
\label{gktr}
\end{equation}%
in which the transformed state $\mathbf{z}$ is realized by the GK state $%
\mathbf{z}^{\prime }\equiv \left( \mathbf{y}^{\prime },\phi ^{\prime
}\right) $ which can be conveniently identified with hybrid (i.e.,
non-canonical) variables. Here $\phi ^{\prime }$ is a suitable
gyrophase-angle (see Section 3) and $\mathbf{y}^{\prime }$ is an appropriate
five-component vector. They are defined in such a way to satisfy either
identically or in asymptotic sense (at least in a suitable subset of
phase-space) the GK equations of motion%
\begin{equation}
\frac{d}{dt}\mathbf{z}^{\prime }=\mathbf{Z},  \label{zeta}
\end{equation}%
in which the vector field $\mathbf{Z}$ is respectively either of the form $%
\mathbf{Z}=\mathbf{Z}\left( \mathbf{y}^{\prime },t\right) $ or $\mathbf{Z}=%
\mathbf{Z}\left( \mathbf{y}^{\prime },t\right) \left[ 1+O\left( \varepsilon
^{k}\right) \right] $, where\ the correction contribution of $O\left(
\varepsilon ^{k}\right) $\textbf{\ }is generally a function of the
gyrophase-angle $\phi ^{\prime }$ too. A fundamental consequence is that the
momentum $p_{\phi ^{\prime }s}$ conjugate to $\phi ^{\prime }$ determines
necessarily one of the phase-functions $K_{j}$. The construction of a
phase-space diffeomorphism of the type (\ref{gktr}) is referred to as GK
theory.

The explicit determination of the phase-function $p_{\phi ^{\prime }s}$
follows from the Hamilton variational functional expressed in hybrid
variables. Starting point is the fundamental differential form in
superabundant Newtonian variables $\left( \mathbf{r},\frac{d}{dt}\mathbf{r},%
\mathbf{v}\right) $ (with $\mathbf{v}$ to be considered independent of $%
\frac{d}{dt}\mathbf{r}$). For a particle characterized by mass $M_{s}$ and
charge $Z_{s}e$ this is given by
\begin{eqnarray}
\Lambda \left( \mathbf{r},\frac{d}{dt}\mathbf{r},\mathbf{v}\right) &\equiv
&dt\mathcal{L}\left( \mathbf{r},\frac{d}{dt}\mathbf{r},\mathbf{v}\right) =
\notag \\
&=&d\mathbf{r}\cdot \left[ M_{s}\mathbf{v}+\frac{Z_{s}e}{c}\mathbf{A}\right]
-dt\mathcal{H}_{s},  \label{questa}
\end{eqnarray}%
where $\mathbf{A}$ is the vector potential, $\mathcal{H}_{s}$ is the total
particle energy%
\begin{equation}
\mathcal{H}_{s}=\frac{M_{s}v^{2}}{2}+Z_{s}e\Phi _{s}^{eff}\equiv {Z_{s}e}%
\Phi _{\ast s}  \label{energii}
\end{equation}%
and $\Phi _{s}^{eff}\equiv \Phi +\frac{M_{s}}{Z_{s}e}\Phi _{G}$, with $\Phi $
and $\Phi _{G}$ being the electrostatic (ES) and gravitational potential
respectively. For definiteness, in the following $\mathbf{A}$ is prescribed
in the Coulomb gauge $\nabla \cdot \mathbf{A}=0$, with both $\mathbf{A}$ and
$\Phi _{s}^{eff}$ to be assumed stationary and $\mathbf{A}$ satisfies the
boundary condition at infinity $\mathbf{A}\rightarrow \mathbf{0}$. By
representing $\Lambda $ in terms of the variables $\mathbf{z}^{\prime }$,
this yields the fundamental Lagrangian differential form $\Lambda _{1}\left(
\mathbf{z}^{\prime },\frac{d}{dt}\mathbf{z}^{\prime }\right) $ of the type%
\begin{eqnarray}
\Lambda _{1}\left( \mathbf{z}^{\prime },\frac{d}{dt}\mathbf{z}^{\prime
}\right) &=&\left[ \Gamma _{y_{i}^{\prime }}+\frac{\partial R}{\partial
y_{i}^{\prime }}\right] dy_{i}^{\prime }+\left[ \Gamma _{\phi ^{\prime }}+%
\frac{\partial R}{\partial \phi ^{\prime }}\right] d\phi ^{\prime }  \notag
\\
&&+\left[ \Gamma _{t}+\frac{\partial R}{\partial t}\right] dt,
\end{eqnarray}%
with $R$ being a generic gauge phase-function. Provided the gauge function $%
R $ and the GK transformation (\ref{gktr}) are suitably defined, $\phi
^{\prime }$ is a GK variable only if it is ignorable for the phase-functions
$\left[ \Gamma _{y_{i}^{\prime }}+\frac{\partial R}{\partial y_{i}^{\prime }}%
\right] $, $\left[ \Gamma _{\phi ^{\prime }}+\frac{\partial R}{\partial \phi
^{\prime }}\right] $ and $\left[ \Gamma _{t}+\frac{\partial R}{\partial t}%
\right] $. This means that $p_{\phi ^{\prime }s}=\left\langle \Gamma _{\phi
^{\prime }}+\frac{\partial R}{\partial \phi ^{\prime }}\right\rangle $,
where the brackets \textquotedblleft $\left\langle {}\right\rangle $%
\textquotedblright\ denote the gyrophase-average performed with respect to
the gyrophase-angle $\phi ^{\prime }$ and keeping constant the remaining GK
variables, namely $\mathbf{y}^{\prime }$. For a generic smooth function of
the GK state this is defined as the linear operator $\left\langle
{}\right\rangle \equiv \frac{1}{2\pi }\int d\phi ^{\prime }$. For
completeness, we introduce here also the tilda-operator $\left[ {}\right]
^{\sim }\equiv I-\left\langle {}\right\rangle $, with $I$ denoting the
identity operator. This permits to define the magnetic moment $m_{s}^{\prime
}$ as the invariant%
\begin{equation}
m_{s}^{\prime }=-\frac{Z_{s}e}{M_{s}c}p_{\phi ^{\prime }s}.  \label{mmm}
\end{equation}%
Notice that in principle the construction of $m_{s}^{\prime }$ can be
formally realized by means of a finite, rather than infinitesimal \cite%
{dubin83,brizard95}, transformation of the type (\ref{gktr}). This task is
carried out in the next section.

\bigskip

\section{Variational non-perturbative GK theory}

In this Section we proceed with the construction of a non-perturbative
approach to non-relativistic GK theory for point-like magnetized charged
particles, namely immersed in stationary and spatially non-symmetric smooth
EM and gravitational fields such that $E^{2}\ll B^{2}$. Starting point is
the definition of the non-canonical variables which prescribe the GK
transformation. In view of the applications to the investigation of kinetic
equilibria we consider here the case\ in which the total particle energy is
conserved, so that $\mathcal{H}_{s}=E_{s}$, where $E_{s}$ is the initial
value of $\mathcal{H}_{s}$. In the following it is also assumed that the
effective potential $\Phi _{s}^{eff}$ is bounded from below, so that, thanks
to energy-conservation, $\left\vert \mathbf{v}\right\vert =\sqrt{\frac{2}{%
M_{s}}\left( E_{s}-Z_{s}e\Phi _{s}^{eff}\right) }$ remains always bounded
too. In this case, a convenient choice for the GK state $\mathbf{z}^{\prime
} $ is to identify it with $\mathbf{z}^{\prime }\equiv \left( \mathbf{r}%
^{\prime },\mathcal{H}_{s},p_{\phi ^{\prime }s},\phi ^{\prime }\right) $,
where $\mathbf{r}^{\prime }$ denotes the guiding-center position vector. As
a consequence of this choice it follows that the vector field $\mathbf{Z}$
introduced in Eq.(\ref{zeta}) takes the form $\mathbf{Z}=\left[ \mathbf{Z}_{%
\mathbf{r}^{\prime }},Z_{\mathcal{H}_{s}}=0,Z_{p_{\phi ^{\prime
}s}}=0,Z_{\phi ^{\prime }}\right] $, with $\mathbf{Z}_{\mathbf{r}^{\prime }}$
and $Z_{\phi ^{\prime }}$ to be determined in terms of the Euler-Lagrange
equations prescribed by the Lagrangian differential form $\Lambda _{1}$.

In order to define the variables $\mathbf{r}^{\prime }$ and $\phi ^{\prime }$
we introduce preliminarily for the Newtonian particle state $\mathbf{x}$ the
transformation to the corresponding guiding-center state $\mathbf{x}^{\prime
}=\left( \mathbf{r}^{\prime },\mathbf{v}^{\prime }\right) $. This is
determined by the guiding-center diffeomorfism:%
\begin{eqnarray}
\mathbf{r} &=&\mathbf{r}^{\prime }+\mathbf{\rho }_{1}^{\prime },
\label{1-bis} \\
\mathbf{v} &=&\mathbf{v}^{\prime }+\mathbf{V}^{\prime },  \label{2-bis}
\end{eqnarray}%
where $\mathbf{r}^{\prime }$ and $\mathbf{v}^{\prime }\equiv u^{\prime }%
\mathbf{b}^{\prime }+\mathbf{w}^{\prime }+\mathbf{\nu }_{1}^{\prime }$ are
respectively referred to as guiding-center position and velocity vectors,
the latter being defined relative to a suitably-prescribed reference
velocity $\mathbf{V}^{\prime }\equiv \mathbf{V}\left( \mathbf{y}^{\prime
},t\right) $. In the following all primed quantities are intended as
evaluated at the position $\mathbf{r}^{\prime }$. Furthermore $\mathbf{\rho }%
_{1}^{\prime }=\mathbf{\rho }_{1}^{\prime }(\mathbf{z}^{\prime },t)$,
denoted as Larmor-radius vector, and the velocity contribution $\mathbf{\nu }%
_{1}^{\prime }=\mathbf{\nu }_{1}^{\prime }(\mathbf{z}^{\prime },t)$ are
assumed to be purely oscillatory, namely of the form $\left\langle \mathbf{%
\rho }_{1}^{\prime }\right\rangle =\left\langle \mathbf{\nu }_{1}^{\prime
}\right\rangle =0$. Instead, $\mathbf{r}^{\prime }$ and $u^{\prime }$ are
assumed to be gyrophase-independent, namely such that $\mathbf{r}^{\prime
}=\left\langle \mathbf{r}^{\prime }\right\rangle $ and $u^{\prime
}=\left\langle u^{\prime }\right\rangle $. In detail, the meaning of the
variables is as follows. First, $\mathbf{b}^{\prime }\equiv \mathbf{b}(%
\mathbf{r}^{\prime },t)$ denotes the unit vector associated to the magnetic
field evaluated at the guiding-center position $\mathbf{r}^{\prime }$. It
follows that $u^{\prime }+\mathbf{\nu }_{1}^{\prime }\cdot \mathbf{b}%
^{\prime }\equiv u^{\prime }+\nu _{1\parallel }^{\prime }$ is the relative
parallel velocity, while $\mathbf{w}^{\prime }+\mathbf{\nu }_{1}^{\prime
}\cdot \left( \underline{\underline{\mathbf{I}}}-\mathbf{b}^{\prime }\mathbf{%
b}^{\prime }\right) $ is the corresponding perpendicular relative velocity.
In particular, $\mathbf{w}^{\prime }$ is represented as $\mathbf{w}^{\prime
}=w^{\prime }\cos \phi ^{\prime }\mathbf{e}_{1}^{\prime }+w^{\prime }\sin
\phi ^{\prime }\mathbf{e}_{2}^{\prime }$, where by construction $w^{\prime
}\equiv \left\vert \mathbf{w}^{\prime }\right\vert =\left\langle w^{\prime
}\right\rangle $, while%
\begin{equation}
\phi ^{\prime }\equiv \arctan \left[ \frac{\left( \mathbf{v}-u^{\prime }%
\mathbf{b}^{\prime }-\mathbf{\nu }_{1}^{\prime }-\mathbf{V}^{\prime }\right)
\cdot \mathbf{e}_{2}^{\prime }}{\left( \mathbf{v}-u^{\prime }\mathbf{b}%
^{\prime }-\mathbf{\nu }_{1}^{\prime }-\mathbf{V}^{\prime }\right) \cdot
\mathbf{e}_{1}^{\prime }}\right]
\end{equation}%
identifies the guiding-center gyrophase angle and $\left( \mathbf{e}%
_{1}^{\prime },\mathbf{e}_{2}^{\prime },\mathbf{e}_{3}^{\prime }=\mathbf{b}%
^{\prime }\right) $ is a right-handed system of orthogonal unit vectors
(magnetic orthonormal basis). Hence, $\phi ^{\prime }$ is a velocity-space
variable which determines the direction of $\mathbf{w}^{\prime }$ in the
plane orthogonal to $\mathbf{b}^{\prime }$ at position $\mathbf{r}^{\prime }$%
. Finally, $\mathbf{V}^{\prime }$ is identified with the
gyrophase-independent species drift velocity%
\begin{equation}
\mathbf{V}^{\prime }=\mathbf{V}_{E^{eff}}^{\prime }+\mathbf{V}_{1}^{\prime
}\equiv V_{\parallel }^{\prime }\mathbf{b}^{\prime }+\mathbf{V}_{\perp
}^{\prime },
\end{equation}%
where $\mathbf{V}_{E^{eff}}^{\prime }\mathbf{=}\frac{c}{B^{\prime }}\mathbf{E%
}^{\prime eff}\times \mathbf{b}^{\prime }$ and $\mathbf{E}^{\prime
eff}\equiv -\nabla ^{\prime }\Phi ^{eff\prime }$ is the species-dependent
effective electric field generated by $\Phi _{s}^{eff\prime }$. In this
representation the species-dependent component $V_{\parallel }^{\prime }$
and the remaining contribution $\mathbf{V}_{1}^{\prime }$ are both to be
considered as suitably-prescribed functions of the GK variables $\mathbf{y}%
^{\prime }$. Due to the choice of GK variables considered here, we need to
establish the relationship between $u^{\prime }$ and $\nu _{1\parallel
}^{\prime }$ in terms of the constant total particle energy $E_{s}$.
Denoting by $D_{s}\left( \mathbf{z}^{\prime }\right) \equiv \sqrt{\frac{2}{%
M_{s}}\left[ E_{s}-Z_{s}e\Phi _{s}^{eff}-\frac{M_{s}}{2}\left( \mathbf{w}%
^{\prime }+\mathbf{\nu }_{1\perp }^{\prime }+\mathbf{V}_{\perp }^{\prime
}\right) ^{2}\right] }$, one obtains the following unique representations:%
\begin{eqnarray}
\left\vert u^{\prime }+V_{\parallel }^{\prime }\right\vert &=&\left\langle
D_{s}\left( \mathbf{z}^{\prime }\right) \right\rangle , \\
\left\vert \nu _{1\parallel }^{\prime }\right\vert &=&\left[ D_{s}\left(
\mathbf{z}^{\prime }\right) \right] ^{\sim }.
\end{eqnarray}%
Notice that the previous quantities demand the evaluation of $\phi ^{\prime
} $-averages in terms of the GK variables $\mathbf{y}^{\prime }$ indicated
above. This requires the corresponding representations for $w^{\prime }$ of
the form $w^{\prime }=w^{\prime }\left( m_{s}^{\prime },E_{s},\mathbf{r}%
^{\prime }\right) $ as well as the explicit determination of $\mathbf{\rho }%
_{1}^{\prime }$, $\mathbf{\nu }_{1}^{\prime }$ and $m_{s}^{\prime }$.

\bigskip

\subsection{GK Lagrangian differential form}

Let us now proceed constructing explicitly the GK Lagrangian differential
form $\Lambda _{1}$ from Eq.(\ref{questa}). Invoking the guiding-center
transformations (\ref{1-bis}) and (\ref{2-bis}), let us introduce the
dynamical gauge $Y=\mathbf{\rho }_{1}^{\prime }\cdot \left( M_{s}\mathbf{V}%
^{\prime }+\frac{Z_{s}e}{c}\mathbf{A}^{\prime }\right) +R\left( \mathbf{z}%
^{\prime }\right) $, with $R\left( \mathbf{z}^{\prime }\right) $ assumed to
be purely oscillatory and to be determined below. Then the Lagrangian
differential form can be equivalent written in terms of $\Lambda _{1}\left(
\mathbf{z}^{\prime },\frac{d}{dt}\mathbf{z}^{\prime }\right) \equiv \Lambda
\left( \mathbf{r},\frac{d}{dt}\mathbf{r},\mathbf{v}\right) -dY$. A
sufficient condition for $\Lambda _{1}$ to be gyrophase-independent is that
the following constraint equations are satisfied identically in the whole
subset of phase-space in which the guiding-center transformation is defined
so that:%
\begin{eqnarray}
\left[ \mathbf{\Gamma }_{\mathbf{r}^{\prime }}-\nabla ^{\prime }R\right]
^{\sim } &=&0,  \label{3} \\
\left[ \Gamma _{\phi ^{\prime }}-\frac{\partial }{\partial \phi ^{\prime }}R%
\right] ^{\sim } &=&0,  \label{4}
\end{eqnarray}%
where $\mathbf{\Gamma }_{\mathbf{r}^{\prime }}=\mathbf{\Gamma }_{\mathbf{r}%
^{\prime }}\left( \mathbf{z}^{\prime }\right) $ and $\Gamma _{\phi ^{\prime
}}=\Gamma _{\phi ^{\prime }}\left( \mathbf{z}^{\prime }\right) $ denote the
phase-functions%
\begin{eqnarray}
\mathbf{\Gamma }_{\mathbf{r}^{\prime }} &\equiv &M_{s}\left( u^{\prime }%
\mathbf{b}^{\prime }+\mathbf{w}^{\prime }+\mathbf{V}^{\prime }+\mathbf{\nu }%
_{1}^{\prime }\right)  \notag \\
&&+\nabla ^{\prime }\mathbf{\rho }_{1}^{\prime }\cdot \left[ M_{s}\left(
u^{\prime }\mathbf{b}^{\prime }+\mathbf{w}^{\prime }+\mathbf{\nu }%
_{1}^{\prime }\right) +\frac{Z_{s}e}{c}\mathbf{A}\right]  \notag \\
&&+\frac{Z_{s}e}{c}\mathbf{A}-\frac{Z_{s}e}{c}\nabla ^{\prime }\mathbf{\rho }%
_{1}^{\prime }\cdot \mathbf{A}^{\prime }  \notag \\
&&-\left[ M_{s}\nabla ^{\prime }\mathbf{V}^{\prime }\cdot \mathbf{\rho }%
_{1}^{\prime }+\frac{Z_{s}e}{c}\nabla ^{\prime }\mathbf{A}^{\prime }\cdot
\mathbf{\rho }_{1}^{\prime }\right] ,  \label{gamma-rprimo}
\end{eqnarray}%
\begin{equation}
\Gamma _{\phi ^{\prime }}\equiv \frac{\partial }{\partial \phi ^{\prime }}%
\mathbf{\rho }_{1}^{\prime }\cdot \left[ M_{s}\left( u^{\prime }\mathbf{b}%
^{\prime }+\mathbf{w}^{\prime }+\mathbf{\nu }_{1}^{\prime }\right) +\frac{%
Z_{s}e}{c}\left( \mathbf{A}-\mathbf{A}^{\prime }\right) \right] .
\label{gamma fi}
\end{equation}%
It follows that, provided a smooth solution to the constraint equations (\ref%
{3}) and (\ref{4}) exists, $\Lambda _{1}\left( \mathbf{z}^{\prime },\frac{d}{%
dt}\mathbf{z}^{\prime }\right) $ becomes%
\begin{equation}
\Lambda _{1}\left( \mathbf{y}^{\prime },\frac{d}{dt}\mathbf{z}^{\prime
}\right) =d\mathbf{r}^{\prime }\cdot \left\langle \mathbf{\Gamma }_{\mathbf{r%
}^{\prime }}-\nabla ^{\prime }R\right\rangle +p_{\phi ^{\prime }s}d\phi
^{\prime }-dt\mathcal{H}_{s},  \label{lambda1}
\end{equation}%
to be referred to as GK Lagrangian differential form. Here the notation is
as follows:%
\begin{eqnarray}
&&\left. \left\langle \mathbf{\Gamma }_{\mathbf{r}^{\prime }}-\nabla
^{\prime }R\right\rangle \equiv M_{s}\left( u^{\prime }\mathbf{b}^{\prime }+%
\mathbf{V}^{\prime }\right) +\frac{Z_{s}e}{c}\left\langle \mathbf{A}%
\right\rangle -\nabla ^{\prime }\left\langle R\right\rangle \right.  \notag
\\
&&\left. +\left\langle \nabla ^{\prime }\mathbf{\rho }_{1}^{\prime }\cdot %
\left[ M_{s}\left( \mathbf{w}^{\prime }+\mathbf{\nu }_{1}^{\prime }\right) +%
\frac{Z_{s}e}{c}\mathbf{A}\right] \right\rangle ,\right.
\end{eqnarray}%
while%
\begin{equation}
p_{\phi ^{\prime }s}\equiv \left\langle \frac{\partial }{\partial \phi
^{\prime }}\mathbf{\rho }_{1}^{\prime }\cdot \left[ M_{s}\left( \mathbf{w}%
^{\prime }+\mathbf{\nu }_{1}^{\prime }\right) +\frac{Z_{s}e}{c}\mathbf{A}%
\right] \right\rangle .  \label{cmmm}
\end{equation}%
The corresponding Euler-Lagrange equations are reported for completeness in
the Appendix. The following remarks are in order:

1) For the extremal variables $\mathbf{r}$ and $\mathbf{v}$ the relationship
$\mathbf{v}=\frac{d}{dt}\mathbf{r}$ is implied by the Hamilton variational
principle. This means that $\mathbf{\rho }_{1}^{\prime }$ and $\mathbf{\nu }%
_{1}^{\prime }$ must necessarily satisfy the additional constraint equation%
\begin{equation}
\frac{d}{dt}\mathbf{\rho }_{1}^{\prime }=\mathbf{w}^{\prime }+\mathbf{\nu }%
_{1}^{\prime },  \label{new}
\end{equation}%
where, due to the chain rule, the lhs becomes%
\begin{equation}
\frac{d}{dt}\mathbf{\rho }_{1}^{\prime }=\mathbf{Z}_{\mathbf{r}^{\prime
}}\cdot \nabla ^{\prime }\mathbf{\rho }_{1}^{\prime }+Z_{\phi ^{\prime }}%
\frac{\partial }{\partial \phi ^{\prime }}\mathbf{\rho }_{1}^{\prime }.
\label{itsnew}
\end{equation}%
Therefore, this implies that the two phase-functions $\mathbf{Z}_{\mathbf{r}%
^{\prime }}$ and $Z_{\phi ^{\prime }}$ enter the very definition of $\mathbf{%
\rho }_{1}^{\prime }$ and $\mathbf{\nu }_{1}^{\prime }$.

2) The evaluation of $\Lambda _{1}$ as well as $p_{\phi ^{\prime }s}$
necessarily involves the preliminary determination of both the vectors $%
\mathbf{\rho }_{1}^{\prime }$ and $\mathbf{\nu }_{1}^{\prime }$ and the
gauge function $R$. This task requires the explicit solution of the set of
constraint equations (\ref{3}), (\ref{4}) and (\ref{new}). The two vector
equations (\ref{3}) and (\ref{new}) determine $\mathbf{\rho }_{1}^{\prime }$
and $\mathbf{\nu }_{1}^{\prime }$, while the scalar equation (\ref{4})
yields the gauge function $R$. This implies that $R$ is a non-local function
of $\mathbf{\rho }_{1}^{\prime }$ and $\mathbf{\nu }_{1}^{\prime }$, while (%
\ref{3}) and (\ref{new}) are non-local equations for the same variables.

3) The non-perturbative expression of $p_{\phi ^{\prime }s}$ in Eq.(\ref%
{cmmm}) recovers the familiar form of a particle canonical momentum. Due to
the constraint equation (\ref{new}) the latter provides for the magnetic
moment the representation%
\begin{eqnarray}
&&\left. m_{s}^{\prime }\equiv -\frac{1}{M_{s}}\left( \frac{Z_{s}e}{c}%
\right) ^{2}\left\langle \frac{\partial }{\partial \phi ^{\prime }}\mathbf{%
\rho }_{1}^{\prime }\cdot \mathbf{A}\right\rangle \right.  \notag \\
&&-\frac{Z_{s}e}{c}\left\langle \frac{\partial }{\partial \phi ^{\prime }}%
\mathbf{\rho }_{1}^{\prime }\cdot \left[ \mathbf{Z}_{\mathbf{r}^{\prime
}}\cdot \nabla ^{\prime }\mathbf{\rho }_{1}^{\prime }+Z_{\phi ^{\prime }}%
\frac{\partial }{\partial \phi ^{\prime }}\mathbf{\rho }_{1}^{\prime }\right]
\right\rangle .  \label{cmmm-bis}
\end{eqnarray}%
Notice that this means in particular that $p_{\phi ^{\prime }s}$ is
generally non-observable. In fact, let us consider the EM gauge
transformation which preserves the conservation of $\mathcal{H}_{s}$, $%
\left( \Phi ,\mathbf{A}\right) \rightarrow \left( \Phi ,\mathbf{A}+\nabla
S\left( \mathbf{r}\right) \right) $, with $S$ denoting a suitable real
smooth scalar function. Then it follows that in such a gauge $p_{\phi
^{\prime }s}$ becomes generally different from Eq.(\ref{cmmm}).

4) For the same reason, also $\mathbf{\rho }_{1}^{\prime }$ and $\mathbf{\nu
}_{1}^{\prime }$ are non-unique and therefore represent physical
non-observable quantities. In fact both the phase-functions $\mathbf{\Gamma }%
_{\mathbf{r}^{\prime }}\left( \mathbf{z}^{\prime }\right) $ and $\Gamma
_{\phi ^{\prime }}\left( \mathbf{z}^{\prime }\right) $ depend explicitly on
the choice of the EM gauge function $S\left( \mathbf{r}\right) $. This means
that also the guiding-center position vector $\mathbf{r}^{\prime }$ as well
as the guiding-center velocity components $u^{\prime }$ and $\mathbf{w}%
^{\prime }$ are generally non-unique (and non-observables). Nevertheless it
is obvious that Eqs.(\ref{1-bis}) and (\ref{2-bis}) warrant that the rhs of
these equations are unique and therefore define observables.

5) To restore the uniqueness of both $p_{\phi ^{\prime }s}$ and the set $%
\left( \mathbf{\rho }_{1}^{\prime },\mathbf{\nu }_{1}^{\prime }\right) $
with respect to the EM gauge transformation indicated above, it is
sufficient to impose that the gauge function $S\left( \mathbf{r}\right) $
vanishes on the boundary of a suitable bounded sub-set $\Omega $ of the
configuration space $%
\mathbb{R}
^{3}$ to which the charged particle belongs. In fact, in this case the
Coulomb gauge $\nabla ^{2}S=0$ requires $S=0$ identically in $\Omega $. This
includes also the case in which $\nabla S$ is a constant vector. Therefore,
under this condition the vector potential $\mathbf{A}$ is uniquely
determined.

6) A fundamental issue is whether under the assumption indicated in point 5)
$p_{\phi ^{\prime }s}$ and the set $\left( \mathbf{\rho }_{1}^{\prime },%
\mathbf{\nu }_{1}^{\prime }\right) $ are actually observables. The answer to
this question is straightforward. In fact, let us consider the action of a
dynamical gauge $R_{1}=R_{1}\left( \mathbf{y}^{\prime }\right) $. This
manifestly leaves invariant the constraint equations (\ref{3}) and (\ref{4})
as well as Eq.(\ref{cmmm}). This proves the statement.

7) We stress that, at this stage, the transformations relating $\mathbf{x}$,
$\mathbf{x}^{\prime }$ and $\mathbf{z}^{\prime }$ are in principle finite
and non-perturbative, which means that they are not based on any
perturbative expansion for the guiding-center transformation and/or the
construction of the GK Lagrangian differential form. The fundamental
physical implication is that, in the subset of phase-space in which the GK
transformation (\ref{gktr}) is defined, provided $\mathbf{\rho }_{1}^{\prime
}$ and $\mathbf{\nu }_{1}^{\prime }$ are uniquely determined, the magnetic
moment $m_{s}^{\prime }$ is a local first-integral of motion.

\bigskip

\section{Second-order LRE for GK theory}

In order to obtain an explicit solution for the constraint equations (\ref{3}%
), (\ref{4}) and (\ref{new}) and to determine explicitly the guiding-center
and GK transformations (respectively, Eqs.(\ref{1-bis})-(\ref{2-bis}) and (%
\ref{gktr})) in this section we develop a second-order perturbative
Larmor-radius expansion (LRE) for all these quantities \cite%
{Cr2011,dubin83,BC85,Bri07,Catto1978,Littlejohn1979,Littlejohn1981,Littke1983,Hahm1988,Balescu,Meiss1990}%
. This is achieved by introducing suitable asymptotic orderings in terms of
the Larmor-radius dimensionless parameter $\varepsilon $ defined as%
\begin{equation}
\varepsilon \equiv \frac{\left\vert \mathbf{\rho }_{1}^{\prime }\right\vert
}{L}\ll 1,  \label{epsilonn}
\end{equation}%
with $L$ denoting the smallest local characteristic scale-length associated
to the stationary EM and gravitational fields. In particular we first impose
the following asymptotic GK orderings:%
\begin{eqnarray}
\left\vert \nabla ^{\prime }\mathbf{\rho }_{1}^{\prime }\right\vert &\sim &%
\frac{\left\vert \mathbf{\nu }_{1}^{\prime }\right\vert }{\left\vert \mathbf{%
v}-\mathbf{\nu }_{1}^{\prime }\right\vert }\sim O\left( \varepsilon \right) ,
\label{ord1} \\
\frac{\left\vert \mathbf{V}^{\prime }\right\vert }{\left\vert \mathbf{v}%
\right\vert } &\lesssim &O\left( \varepsilon ^{0}\right) ,  \label{ord2} \\
\frac{\frac{M_{s}v^{2}}{2}}{Z_{s}e\Phi _{s}^{eff}} &\sim &\frac{%
M_{s}\left\vert \mathbf{v}\right\vert }{\frac{Z_{s}e}{c}\left\vert \mathbf{A}%
\right\vert }\sim O\left( \varepsilon \right) ,  \label{ord3}
\end{eqnarray}%
and furthermore for the two fields $\mathbf{Z}_{\mathbf{r}^{\prime }}$ and $%
Z_{\phi ^{\prime }}$ we require%
\begin{eqnarray}
\frac{\left\vert \mathbf{Z}_{\mathbf{r}^{\prime }}\right\vert }{L\Omega
_{cs}^{\prime }} &\sim &O\left( \varepsilon \right) , \\
\left\vert \frac{Z_{\phi ^{\prime }}}{\Omega _{cs}^{\prime }}\right\vert
&=&1+O\left( \varepsilon \right) ,
\end{eqnarray}%
where $\Omega _{cs}^{\prime }\equiv \frac{Z_{s}eB^{\prime }}{M_{s}c}$ is the
cyclotron frequency at the guiding-center position $\mathbf{r}^{\prime }$.
Then, introducing for $\mathbf{\rho }_{1}^{\prime }$ and $\mathbf{\nu }%
_{1}^{\prime }$ perturbative expansions of the form%
\begin{eqnarray}
\mathbf{\rho }_{1}^{\prime } &=&\varepsilon \mathbf{r}_{1}^{\prime
}+\varepsilon ^{2}\mathbf{r}_{2}^{\prime }+..., \\
\mathbf{\nu }_{1}^{\prime } &=&\varepsilon \mathbf{v}_{1}^{\prime
}+\varepsilon ^{2}\mathbf{v}_{2}^{\prime }+...,
\end{eqnarray}%
the second-order guiding-center transformation is thus obtained by letting%
\begin{eqnarray}
\mathbf{r} &=&\mathbf{r}^{\prime }+\varepsilon \mathbf{r}_{1}^{\prime
}+\varepsilon ^{2}\mathbf{r}_{2}^{\prime },  \label{neve-1} \\
\mathbf{v} &=&u^{\prime }\mathbf{b}^{\prime }+\mathbf{w}^{\prime }+\mathbf{V}%
^{\prime }+\varepsilon \mathbf{v}_{1}^{\prime },
\end{eqnarray}%
while the gauge function $R$ must be similarly expanded as $R=\varepsilon
R_{1}+\varepsilon ^{2}R_{2}+...$ Let us now assume that the EM and
gravitational fields are smooth-$C^{\infty }$ functions which vary slowly on
the Larmor-radius scale, in the sense that $\mathbf{A}=\mathbf{A}%
(\varepsilon ^{k}\mathbf{r})$ and $\Phi ^{eff}=\Phi ^{eff}(\varepsilon ^{k}%
\mathbf{r})$, with $k\geq 0$. In the case $k=0$, this implies for them the
validity of Taylor-expansions of the form%
\begin{eqnarray}
\mathbf{A}(\mathbf{r}) &=&\mathbf{A}^{\prime }+\varepsilon \mathbf{r}%
_{1}^{\prime }\cdot \nabla ^{\prime }\mathbf{A}^{\prime }+\varepsilon ^{2}%
\frac{1}{2}\mathbf{r}_{1}^{\prime }\mathbf{r}_{1}^{\prime }:\nabla ^{\prime
}\nabla ^{\prime }\mathbf{A}^{\prime }  \notag \\
&&+\varepsilon ^{2}\mathbf{r}_{2}^{\prime }\cdot \nabla ^{\prime }\mathbf{A}%
^{\prime }\left[ 1+O\left( \varepsilon \right) \right] , \\
\Phi ^{eff}(\mathbf{r}) &=&\Phi ^{eff\prime }+\varepsilon \mathbf{r}%
_{1}^{\prime }\cdot \nabla ^{\prime }\Phi ^{eff\prime }+\varepsilon ^{2}%
\frac{1}{2}\mathbf{r}_{1}^{\prime }\mathbf{r}_{1}^{\prime }:\nabla ^{\prime
}\nabla ^{\prime }\Phi ^{eff\prime }  \notag \\
&&+\varepsilon ^{2}\mathbf{r}_{2}^{\prime }\cdot \nabla ^{\prime }\Phi
^{eff\prime }\left[ 1+O\left( \varepsilon \right) \right] .
\end{eqnarray}%
Introducing the previous representations in the constraint equations (\ref{3}%
), (\ref{4}) and (\ref{new}) one readily finds that, correct to first order
in $\varepsilon $, $\mathbf{r}_{1}^{\prime }$ is given by the customary
expression%
\begin{equation}
\mathbf{r}_{1}^{\prime }=-\frac{\mathbf{w}^{\prime }\times \mathbf{b}%
^{\prime }}{\Omega _{cs}^{\prime }}.  \label{r1dirif}
\end{equation}%
Instead, $\mathbf{r}_{2}^{\prime }$ is found to obey the constraint equation%
\begin{eqnarray}
&&M_{s}\mathbf{v}_{1}^{\prime }+\frac{Z_{s}e}{c}\left[ \frac{1}{2}\mathbf{r}%
_{1}^{\prime }\mathbf{r}_{1}^{\prime }:\nabla ^{\prime }\nabla ^{\prime }%
\mathbf{A}^{\prime }\right] ^{\sim }  \notag \\
&&+\left[ \nabla ^{\prime }\mathbf{r}_{1}^{\prime }\cdot \left[ M_{s}\left(
u^{\prime }\mathbf{b}^{\prime }+\mathbf{w}^{\prime }\right) +\frac{Z_{s}e}{c}%
\mathbf{r}_{1}^{\prime }\cdot \nabla ^{\prime }\mathbf{A}^{\prime }\right] %
\right] ^{\sim }  \notag \\
&&\left. -\frac{Z_{s}eB^{\prime }}{c}\mathbf{r}_{2}^{\prime }\times \mathbf{b%
}^{\prime }-M_{s}\nabla ^{\prime }\mathbf{V}^{\prime }\cdot \mathbf{r}%
_{1}^{\prime }=\left[ \nabla ^{\prime }R_{1}\right] ^{\sim },\right.
\label{v1r2-gg}
\end{eqnarray}%
with the gauge function $R_{1}$ being given by%
\begin{eqnarray}
R_{1} &=&\frac{Z_{s}e}{c\Omega _{c}^{\prime }}\int d\phi ^{\prime }\left[
\left( \mathbf{r}_{1}^{\prime }\cdot \nabla ^{\prime }\mathbf{A}^{\prime
}\right) \cdot \mathbf{w}^{\prime }\right] ^{\sim }  \notag \\
&&-\left\langle \frac{Z_{s}e}{c\Omega _{c}^{\prime }}\int d\phi ^{\prime }%
\left[ \left( \mathbf{r}_{1}^{\prime }\cdot \nabla ^{\prime }\mathbf{A}%
^{\prime }\right) \cdot \mathbf{w}^{\prime }\right] ^{\sim }\right\rangle .
\end{eqnarray}

As a consequence, the magnetic moment representation given by Eq.(\ref%
{cmmm-bis}) yields correct to first order in $\varepsilon $ the expression%
\begin{equation}
m_{s}^{\prime }=\mu _{0}^{\prime }+\varepsilon \mu _{1}^{\prime },
\label{cipolla}
\end{equation}%
where $\mu _{0}^{\prime }\equiv \frac{1}{2}\frac{M_{s}}{B^{\prime }}%
w^{\prime 2}$ is the customary representation for the leading-order magnetic
moment, while the first-order correction is given by%
\begin{equation}
\varepsilon \mu _{1}^{\prime }\equiv \varepsilon \frac{M_{s}}{B^{\prime }}%
\left\langle \mathbf{w}^{\prime }\cdot \mathbf{v}_{1}^{\prime }\right\rangle
.
\end{equation}%
In particular, the scalar product $\left\langle \mathbf{w}^{\prime }\cdot
\mathbf{v}_{1}^{\prime }\right\rangle $ can be evaluated explicitly based on
Eq.(\ref{v1r2-gg}) yielding%
\begin{equation}
\left\langle \mathbf{w}^{\prime }\cdot \mathbf{v}_{1}^{\prime }\right\rangle
=-\left\langle \mathbf{w}^{\prime }\cdot \nabla ^{\prime }\left( u^{\prime }%
\mathbf{b}^{\prime }+\mathbf{V}^{\prime }\right) \cdot \mathbf{r}%
_{1}^{\prime }\right\rangle +\Omega _{cs}^{\prime 2}\left\langle \mathbf{r}%
_{2}^{\prime }\cdot \mathbf{r}_{1}^{\prime }\right\rangle ,  \label{stul}
\end{equation}%
where in particular%
\begin{eqnarray}
&&\left. \left\langle \mathbf{w}^{\prime }\cdot \nabla ^{\prime }\left(
u^{\prime }\mathbf{b}^{\prime }+\mathbf{V}^{\prime }\right) \cdot \mathbf{r}%
_{1}^{\prime }\right\rangle =\right.  \notag \\
&&\frac{c}{B^{\prime }Z_{s}e}\mu ^{\prime }\left[ u\frac{4\pi }{c}%
J_{\parallel }^{\prime }+\mathbf{B}^{\prime }\cdot \nabla ^{\prime }\times
\mathbf{V}^{\prime }\right] ,  \label{pan}
\end{eqnarray}%
with $J_{\parallel }^{\prime }\equiv \mathbf{J}_{\parallel }^{\prime }\cdot
\mathbf{b}^{\prime }$ being the parallel electric current density which
locally generates the non-vacuum magnetic field. From Eq.(\ref{stul}) it
follows that, in order to determine the correction $\varepsilon \mu
_{1}^{\prime }$, the explicit representation of $\mathbf{r}_{2}^{\prime }$
is required. In order to proceed with this, we introduce the following
perturbative expansions for the two quantities $\mathbf{Z}_{\mathbf{r}%
^{\prime }}$ and $Z_{\phi ^{\prime }}$ which enter the constraint equation (%
\ref{itsnew}):%
\begin{eqnarray}
\mathbf{Z}_{\mathbf{r}^{\prime }} &=&\left[ \mathbf{Z}_{\mathbf{r}^{\prime }}%
\right] _{0}+\varepsilon \left[ \mathbf{Z}_{\mathbf{r}^{\prime }}\right]
_{1},  \label{zidle-1} \\
Z_{\phi ^{\prime }} &=&\frac{1}{\varepsilon }\left[ Z_{\phi ^{\prime }}%
\right] _{-1}+\left[ Z_{\phi ^{\prime }}\right] _{0}.  \label{zidle-2}
\end{eqnarray}%
In particular, from the Euler-Lagrange equations given in the Appendix,
after setting $V_{\parallel }^{\prime }=0$, one can obtain for the terms in
Eq.(\ref{zidle-1}) the following result%
\begin{eqnarray}
\left[ \mathbf{Z}_{\mathbf{r}^{\prime }}\right] _{0} &=&u^{\prime }\mathbf{b}%
^{\prime }+\mathbf{V}^{\prime }, \\
\varepsilon \left[ \mathbf{Z}_{\mathbf{r}^{\prime }}\right] _{1} &=&\frac{%
\varepsilon u^{\prime }}{\Omega _{cs}^{\prime }}\left( u^{\prime }\frac{4\pi
}{cB^{\prime }}\mathbf{J}^{\prime }+\nabla ^{\prime }\times \mathbf{V}%
^{\prime }\right) \cdot \left( \underline{\mathbf{I}}-\mathbf{bb}\right)
\notag \\
&&-\frac{\varepsilon }{\Omega _{cs}^{\prime }}\left[ u^{\prime 2}\nabla
^{\prime }\ln B^{\prime }\times \mathbf{b}^{\prime }+\frac{\mu ^{\prime }}{%
M_{s}}\nabla ^{\prime }B^{\prime }\right] \times \mathbf{b}^{\prime }  \notag
\\
&&-\frac{\varepsilon }{\Omega _{cs}^{\prime }}\left[ \nabla \mathbf{V}%
^{\prime }\cdot \mathbf{V}^{\prime }\right] \times \mathbf{b}^{\prime },
\end{eqnarray}%
where in these expressions and in the following calculations $\mathbf{V}%
^{\prime }=\mathbf{V}_{E^{eff}}^{\prime }$. Similarly, the series terms in
Eq.(\ref{zidle-2}) are found to be%
\begin{eqnarray}
\left[ Z_{\phi ^{\prime }}\right] _{-1} &=&-\Omega _{cs}^{\prime }, \\
\left[ Z_{\phi ^{\prime }}\right] _{0} &=&\frac{\Omega _{cs}^{\prime }}{%
w^{\prime 2}}\left\langle \mathbf{w}^{\prime }\cdot \nabla ^{\prime }\left(
u^{\prime }\mathbf{b}^{\prime }+\mathbf{V}^{\prime }\right) \cdot \mathbf{r}%
_{1}^{\prime }\right\rangle  \notag \\
&&+\frac{\Omega _{cs}^{\prime }}{w^{\prime 2}}\left\langle \mathbf{Z}_{%
\mathbf{r}^{\prime }0}\cdot \nabla ^{\prime }\mathbf{r}_{1}^{\prime }\cdot
\mathbf{w}^{\prime }\right\rangle .
\end{eqnarray}%
Then, combining Eqs.(\ref{new}) and (\ref{v1r2-gg}), the following equation
is obtained for $\mathbf{r}_{2}^{\prime }$:%
\begin{equation}
\frac{\partial }{\partial \phi ^{\prime }}\mathbf{r}_{2}^{\prime }+\mathbf{r}%
_{2}^{\prime }\times \mathbf{b}^{\prime }=\mathbf{S}_{2}^{\sim },
\label{eqr222}
\end{equation}%
where%
\begin{eqnarray}
\mathbf{S}_{2}^{\sim } &=&\frac{1}{\Omega _{cs}^{\prime }}\left[ \left[
\mathbf{Z}_{\mathbf{r}^{\prime }}\right] _{0}\cdot \nabla ^{\prime }\mathbf{r%
}_{1}^{\prime }+\left[ Z_{\phi ^{\prime }}\right] _{0}\frac{\partial }{%
\partial \phi ^{\prime }}\mathbf{r}_{1}^{\prime }\right] -\frac{\left[
\nabla ^{\prime }R_{1}\right] ^{\sim }}{M_{s}\Omega _{cs}^{\prime }}  \notag
\\
&&+\frac{1}{\Omega _{cs}^{\prime }}\left[ \nabla ^{\prime }\mathbf{r}%
_{1}^{\prime }\cdot \left[ \left( u^{\prime }\mathbf{b}^{\prime }+\mathbf{w}%
^{\prime }\right) +\frac{Z_{s}e}{cM_{s}}\mathbf{r}_{1}^{\prime }\cdot \nabla
^{\prime }\mathbf{A}^{\prime }\right] \right] ^{\sim }  \notag \\
&&+B^{\prime }\left[ \frac{1}{2}\mathbf{r}_{1}^{\prime }\mathbf{r}%
_{1}^{\prime }:\nabla ^{\prime }\nabla ^{\prime }\mathbf{A}^{\prime }\right]
^{\sim }-\frac{1}{\Omega _{cs}^{\prime }}\nabla ^{\prime }\mathbf{V}^{\prime
}\cdot \mathbf{r}_{1}^{\prime }.  \label{esse-2}
\end{eqnarray}%
In order to determine an explicit solution for $\mathbf{r}_{2}^{\prime }$,
we project Eq.(\ref{eqr222}) along the magnetic orthonormal basis $\left(
\mathbf{e}_{1}^{\prime },\mathbf{e}_{2}^{\prime },\mathbf{e}_{3}^{\prime }=%
\mathbf{b}^{\prime }\right) $. The parallel component determines $%
r_{2\parallel }^{\prime }\equiv \mathbf{r}_{2}^{\prime }\cdot \mathbf{b}%
^{\prime }$, yielding%
\begin{equation}
r_{2\parallel }^{\prime }=\int d\phi ^{\prime }\left[ \mathbf{S}_{2}^{\sim
}\cdot \mathbf{b}^{\prime }\right] -\left\langle \int d\phi ^{\prime }\left[
\mathbf{S}_{2}^{\sim }\cdot \mathbf{b}^{\prime }\right] \right\rangle .
\end{equation}%
The projections along $\mathbf{e}_{1}^{\prime }$ and $\mathbf{e}_{2}^{\prime
}$ can be combined to give the set of equations%
\begin{eqnarray}
\frac{\partial ^{2}}{\partial \phi ^{\prime 2}}r_{21}^{\prime
}+r_{21}^{\prime } &=&D_{1}\left( \phi ^{\prime }\right) ,
\label{OSC-HARM-1} \\
\frac{\partial ^{2}}{\partial \phi ^{\prime 2}}r_{22}^{\prime
}+r_{22}^{\prime } &=&D_{2}\left( \phi ^{\prime }\right) ,
\label{OSC-HARM-2}
\end{eqnarray}%
where $r_{21}^{\prime }\equiv \mathbf{r}_{2}^{\prime }\cdot \mathbf{e}%
_{1}^{\prime }$ and $r_{22}^{\prime }\equiv \mathbf{r}_{2}^{\prime }\cdot
\mathbf{e}_{2}^{\prime }$, and furthermore%
\begin{eqnarray}
D_{1}\left( \phi ^{\prime }\right) &\equiv &\frac{\partial }{\partial \phi
^{\prime }}\mathbf{S}_{2}^{\sim }\cdot \mathbf{e}_{1}^{\prime }-\mathbf{S}%
_{2}^{\sim }\cdot \mathbf{e}_{2}^{\prime }, \\
D_{2}\left( \phi ^{\prime }\right) &\equiv &\frac{\partial }{\partial \phi
^{\prime }}\mathbf{S}_{2}^{\sim }\cdot \mathbf{e}_{2}^{\prime }+\mathbf{S}%
_{2}^{\sim }\cdot \mathbf{e}_{1}^{\prime }.
\end{eqnarray}%
In both cases, the general solution can be written as%
\begin{eqnarray}
r_{21}^{\prime } &=&\alpha r_{11}^{\prime }+\int_{0}^{2\pi }d\overline{\phi }%
\digamma \left( \phi ^{\prime },\overline{\phi }\right) D_{1}\left(
\overline{\phi }\right) , \\
r_{22}^{\prime } &=&\beta r_{12}^{\prime }+\int_{0}^{2\pi }d\overline{\phi }%
\digamma \left( \phi ^{\prime },\overline{\phi }\right) D_{2}\left(
\overline{\phi }\right) ,
\end{eqnarray}%
where $r_{11}^{\prime }\equiv \mathbf{r}_{1}^{\prime }\cdot \mathbf{e}%
_{1}^{\prime }$ and $r_{12}^{\prime }\equiv \mathbf{r}_{1}^{\prime }\cdot
\mathbf{e}_{2}^{\prime }$, $\digamma \left( \phi ^{\prime },\overline{\phi }%
\right) =\sin \left( \phi ^{\prime }-\overline{\phi }\right) $ is the Green
function for Eqs.(\ref{OSC-HARM-1}) and (\ref{OSC-HARM-2}) and $\alpha
,\beta $ are gyrophase-independent. In the context of the present
perturbative theory, the terms $\alpha r_{11}^{\prime }$ and $\alpha \beta
r_{12}^{\prime }$ represent corrections proportional to $\mathbf{r}%
_{1}^{\prime }$. Therefore, the coefficients $\alpha $ and $\beta $ can
always be set equal to zero. The explicit determination of the two
components $r_{21}^{\prime }$ and $r_{22}^{\prime }$ can be readily obtained
based on Eq.(\ref{esse-2}). Combining Eqs.(\ref{stul}) and (\ref{pan}) and
the result for $\mathbf{r}_{2}^{\prime }$ yields the following expression
for the first-order perturbation $\varepsilon \mu _{1}^{\prime }$:%
\begin{eqnarray}
\varepsilon \mu _{1}^{\prime } &=&-\varepsilon \mu _{0}^{\prime }\frac{1}{%
B^{\prime }}\frac{1}{\Omega _{cs}^{\prime }}\left[ u^{\prime }\frac{4\pi }{c}%
J_{\parallel }^{\prime }+\mathbf{B}^{\prime }\cdot \nabla ^{\prime }\times
\mathbf{V}^{\prime }\right]  \notag \\
&&-\varepsilon \mu _{0}^{\prime }\frac{2}{\Omega _{cs}^{\prime }}\left[
\nabla ^{\prime }\cdot \mathbf{V}^{\prime }+\frac{4\pi }{B^{\prime 2}}%
\mathbf{J}^{\prime }\cdot \mathbf{E}^{\prime eff}\right]  \notag \\
&&+\varepsilon \mu _{0}^{\prime }\frac{1}{\Omega _{cs}^{\prime }}\frac{8\pi
}{B^{\prime 2}}J_{\parallel }^{\prime }E_{\parallel }^{\prime eff},
\label{mu11}
\end{eqnarray}%
where $E_{\parallel }^{\prime eff}\equiv \mathbf{E}^{\prime eff}\cdot
\mathbf{b}^{\prime }$.

Let us briefly analyze the physical interpretation of Eq.(\ref{mu11}). The
first two terms on the rhs formally agree with the perturbative calculations
obtained in Ref.\cite{BC85} for the magnetic moment evaluated at the actual
particle position $\mathbf{r}$. The remaining contributions on the rhs are
new and represent corrections arising due to the presence of both a finite
current density $\mathbf{J}^{\prime }$ and a strong effective electric
field, in the sense of Eqs.(\ref{ord2}) and (\ref{ord3}). In more detail, we
notice that the first contribution on the rhs which is proportional to $%
u^{\prime }$ coincides with the term identified originally by Kruskal \cite%
{kruskal65} and ascribed to the existence of a non-vanishing parallel
current density $J_{\parallel }^{\prime }$. Remarkably, at this order, the
latter is the only contribution to the magnetic moment which is odd in the
parallel velocity $u^{\prime }$.

We conclude\ this section pointing out that the existence of additional
adiabatic invariants should be excluded in the present context. In
particular, the parallel adiabatic invariant must be generally ruled out
because of the ordering assumption (\ref{ord3}) invoked for the effective
potential together with the absence of symmetry for the EM and gravitational
fields. Therefore, under these conditions and in validity of the GK
transformation and the LRE, it follows that the only independent adiabatic
invariants which characterize the system of interest are the total particle
effective potential $\Phi _{\ast s}$ defined in Eq.(\ref{energii}) and the
magnetic moment $m_{s}^{\prime }$ given by Eqs.(\ref{cipolla}) and (\ref%
{mu11}). For definiteness, in the following we consider the subset of
phase-space for which the adiabatic invariants $G_{s}=\left( \Phi _{\ast
s},m_{s}^{\prime }\right) $ exist and are assumed such that $\frac{1}{\Omega
_{s}^{\prime }}\frac{d}{dt}\ln G_{s}=0+O\left( \varepsilon _{s}^{k}\right) $%
, with $k\geq 3$, while possible lower-order adiabatic invariants will be
ignored.

\bigskip

\section{Kinetic equilibria: general solution}

In this section we address the issue of the construction of Vlasov-Maxwell
spatially non-symmetric kinetic equilibria, which describe collisionless
plasmas immersed in spatially non-symmetric and smooth EM and gravitational
fields. As stated in the Introduction, here we focus on the treatment of
flow-through plasma equilibria occurring in suitable sub-sets of the
configuration space and in the presence of appropriate plasma sources. In
particular, we look for solutions of the Vlasov equation in the sub-set of
velocity space $W\subseteq
\mathbb{R}
^{3}$ in which the magnetic moment\ $m_{s}^{\prime }$ is an adiabatic
invariant. We restrict our analysis to the case of slowly-varying EM and
gravitational fields, ignoring the possible occurrence of both global and
local (in velocity space) resonance phenomena for the magnetic moment \cite%
{ciri,fi1,fi2,fi3}. Following the corresponding treatment developed in Refs.%
\cite{Cr2010,Cr2011,Cr2011a,Cr2012} for axisymmetric kinetic equilibria, the
solutions are taken of the form%
\begin{equation}
f_{\ast s}=f_{\ast s}\left( \Phi _{\ast s},m_{s}^{\prime },\Lambda _{\ast
s},\lambda ^{k}t\right) ,  \label{form}
\end{equation}%
with $k\geq 1$ and $\lambda $ being a suitable small dimensionless parameter
to be later identified, so that $f_{\ast s}$ is at most slowly explicitly
dependent on $t$. Here $f_{\ast s}$ is assumed to be a strictly-positive
real function which depends on the set $G_{s}=\left( \Phi _{\ast
s},m_{s}^{\prime }\right) $ both explicitly and implicitly. In addition, $%
f_{\ast s}$ is summable, in the sense that the velocity moments of the form $%
\Xi _{s}\left( \mathbf{r}\right) =\int_{W}d^{3}vX_{s}\left( \mathbf{x}%
\right) f_{\ast s}$ must exist for a suitable ensemble of weight functions $%
\left\{ X_{s}\left( \mathbf{x}\right) \right\} $, to be prescribed in terms
of polynomials of arbitrary degree defined with respect to components of the
velocity vector field $\mathbf{v}$. The implicit dependences in $f_{\ast s}$
occur through analytic functions $\Lambda _{\ast s}$, which identify the
so-called structure functions. Hence, $\Lambda _{\ast s}$ must be a function
of the same invariants $G_{s}$. The latter restriction is referred to as a
kinetic constraint and is expressed by imposing that%
\begin{equation}
\Lambda _{\ast s}=\Lambda _{\ast s}\left( \Phi _{\ast s},m_{s}^{\prime
}\right) .  \label{lamba}
\end{equation}%
As discussed in the next section, under suitable assumptions the choice of
the functions $\Lambda _{\ast s}$ is shown to be related to the construction
of an appropriate set of observable associated with the same KDF. A
particular realization for the kinetic constraint (\ref{lamba}) is%
\begin{equation}
\Lambda _{\ast s}=\Lambda _{\ast s}\left( \Phi _{\ast s},\lambda
^{j}m_{s}^{\prime }\right) ,
\end{equation}%
with $j\geq 1$. In this case the structure functions $\Lambda _{\ast s}$
admit the series expansion%
\begin{equation}
\Lambda _{\ast s}=\Lambda _{\ast s}^{\left( 0\right) }\left( \Phi _{\ast
s}\right) +\lambda ^{j}\Lambda _{\ast s}^{\left( 1\right) }\left( \Phi
_{\ast s},m_{s}^{\prime }\right) .  \label{espan-1}
\end{equation}

In view of the theory developed in Refs.\cite{Cr2010,Cr2011,Cr2011a,Cr2012}
and the previous considerations, a convenient choice for $f_{\ast s}$ is
represented by the \textit{generalized bi-Maxwellian distribution }expressed
as%
\begin{equation}
f_{\ast s}=\frac{\beta _{\ast s}}{\left( 2\pi /M_{s}\right) ^{3/2}\left(
T_{\parallel \ast s}\right) ^{1/2}}\exp \left\{ -\frac{{Z_{s}e}\Phi _{\ast s}%
}{T_{\parallel \ast s}}-m_{s}^{\prime }\alpha _{\ast s}\right\} .
\label{sol2}
\end{equation}%
Here $\alpha _{\ast s}\equiv \frac{B}{\Delta _{T_{s\ast }}}$, the quantities
$\frac{1}{\Delta _{T_{s}\ast }}\equiv \frac{1}{T_{\perp s}}-\frac{1}{%
T_{\parallel \ast s}}$, $T_{\perp s}$, $T_{\parallel \ast s}$ are denoted
respectively as generalized species temperature anisotropy, perpendicular
and parallel temperatures, while $\beta _{\ast s}\equiv \frac{\eta _{s}}{%
T_{\perp s}}$, with $\eta _{s}$ to be referred to as the generalized species
pseudo-density. In Eq.(\ref{sol2}) the structure functions are identified
with the set $\left\{ \Lambda _{\ast s}\right\} \equiv \left\{ \beta _{\ast
s},\alpha _{\ast s},T_{\parallel \ast s}\right\} $. Notice that, at this
stage, the functional form of the structure functions remains still
arbitrary, so that they cannot be directly identified with particular fluid
fields, i.e. velocity moments of the KDF. This means that the KDF $f_{\ast
s} $ in Eq.(\ref{sol2}) is always an exact solution of the Vlasov equation,
but in a strict sense it remains non-Gaussian in velocity space, although
its asymptotic behavior in velocity space must be such to warrant the
existence of the velocity moments $\left\{ \Xi _{s}\left( \mathbf{r}\right)
\right\} $.

\bigskip

\section{Perturbative theory: Maxwellian-like KDFs}

In this section we prove that the general solution for the species
equilibrium KDF given by Eq.(\ref{sol2}) admits an asymptotic approximation
which corresponds to a Maxwellian-like equilibrium and we investigate the
physical conditions under which this happens. In particular, here we search
for the asymptotic orderings which warrant that $f_{\ast s}$ in Eq.(\ref%
{sol2}) becomes asymptotically \textquotedblleft close\textquotedblright\
(in a sense to be specified below) to a local bi-Maxwellian KDF, at least in
a suitable sub-set of phase-space. Following Refs.\cite%
{Cr2010,Cr2011,Cr2011a,Cr2012}, this goal is reached by developing a
perturbative theory which permits to treat explicitly the implicit
functional dependences on the particle invariants carried by the structure
functions in Eq.(\ref{sol2}).

In detail, we restrict here the analysis to the thermal sub-set of the
velocity sub-space $W$ such that the following ordering assumption holds:%
\begin{equation}
\frac{\left\vert \mathbf{v}\right\vert }{v_{ths}}\sim O\left( 1\right) ,
\label{thermall}
\end{equation}%
where $v_{ths}\equiv \sqrt{2T_{s}/M_{s}}$ is the\ species thermal velocity
defined with respect to the plasma species temperature $T_{s}$. In such a
set the Larmor radius $r_{Ls}$ is of order $r_{Ls}\sim v_{ths}/\Omega _{cs}$%
, so that the Larmor-radius parameter can be identified with $\varepsilon
_{s}\equiv \frac{v_{ths}}{\Omega _{cs}L}\ll 1$. Similarly, for thermal
particles satisfying the ordering (\ref{thermall}) we introduce the
independent parameter $\sigma _{s}\equiv \left\vert \frac{T_{s}}{{Z_{s}e}%
\Phi _{s}^{eff}}\right\vert $ which determines the ratio between thermal
kinetic energy and the particle effective potential energy. Adopting the
classification of kinetic regimes for collisionless plasmas developed in Ref.%
\cite{Cr2012}, we shall assume that the species thermal particles belong to
the \emph{strong effective potential energy regime (SEPE regime)}, namely
the sub-set of\ the one-particle ($s$-species)\ phase-space for which the
asymptotic ordering $\sigma _{s}\ll 1$ holds. In this case the particle
kinetic energy $\frac{M_{s}v^{2}}{2}$ is ordered as $O\left( \sigma
_{s}\right) $ with respect to the effective potential $\Phi _{s}^{eff}$, so
that all smooth functions of $\Phi _{\ast s}$ can be Taylor-expanded around $%
\Phi _{s}^{eff}$.

Invoking now the $\sigma _{s}$-ordering appropriate for thermal particles in
the SEPE regime and the series representation (\ref{cipolla}) for $%
m_{s}^{\prime }$, after identifying the parameter $\lambda $ with $%
\varepsilon _{s}$ in Eq.(\ref{espan-1}), correct to first order in $\sigma
_{s}$ and ignoring terms of $O\left( \sigma _{s}\right) O\left( \varepsilon
_{s}^{j}\right) $ with $j\geq 1$, one obtains%
\begin{eqnarray}
\Lambda _{\ast s} &=&\Lambda _{s}^{\left( 0\right) }\left( \Phi
_{s}^{eff}\right) +\varepsilon _{s}^{j}\Lambda _{s}^{\left( 1\right) }\left(
\Phi _{s}^{eff},\mu _{0s}^{\prime }\right)  \notag \\
&&+\sigma _{s}\left( \Phi _{\ast s}-\Phi _{s}^{eff}\right) \frac{\partial
\Lambda _{s}^{\left( 0\right) }}{\partial \Phi _{s}^{eff}}.
\label{taylor lambda}
\end{eqnarray}%
In terms of Eq.(\ref{taylor lambda}) a Chapman-Enskog representation can be
recovered for $f_{\ast s}$, yielding%
\begin{equation}
f_{\ast s}=f_{bi-M,s}^{\prime }\left[ 1+\sigma _{s}h_{s}^{\left( 1\right)
}+\varepsilon _{s}^{j}h_{s}^{\left( 2\right) }\right] ,  \label{chens}
\end{equation}%
where $f_{bi-M,s}^{\prime }\equiv f_{bi-M,s}^{\prime }\left( \Phi _{\ast
s},m_{s}^{\prime },\Lambda _{s}^{\left( 0\right) },\lambda ^{k}t\right) $,
while $h_{s}^{\left( 1\right) }$ and $h_{s}^{\left( 2\right) }$ represent
respectively the perturbative corrections carried by the $\sigma _{s}$- and $%
\varepsilon _{s}$-expansions of the structure functions $\Lambda _{\ast s}$.
Plasma species satisfying Eq.(\ref{chens}) will be referred to below as
local-Maxwellian species. In this case the leading-order KDF $%
f_{bi-M,s}^{\prime }$ is found to be%
\begin{equation}
f_{bi-M,s}^{\prime }=\frac{\beta _{s}\exp \left\{ -\frac{{Z_{s}e}\Phi
_{s}^{eff}}{T_{\parallel s}}\right\} }{\left( 2\pi /M_{s}\right)
^{3/2}T_{\parallel s}^{1/2}}\exp \left\{ -\frac{M_{s}v^{2}}{2T_{\parallel s}}%
-m_{s}^{\prime }\alpha _{s}\right\} ,  \label{fbima}
\end{equation}%
where now the leading-order structure functions are only position-dependent
and are identified with the set $\left\{ \Lambda _{s}^{\left( 0\right)
}\right\} =\left\{ \beta _{s},\alpha _{s},T_{\parallel s}\right\} $ which
depends functionally only on $\Phi _{s}^{eff}$. More precisely, here $\alpha
_{s}\equiv \frac{B}{\Delta _{T_{s}}}$, with $\frac{1}{\Delta _{T_{s}}}\equiv
\frac{1}{T_{\perp s}}-\frac{1}{T_{\parallel s}}$ being the species
temperature anisotropy, where $T_{\perp s}$ and $T_{\parallel s}$ are
respectively the leading-order species perpendicular and parallel
temperatures. In addition, $\beta _{s}\equiv \frac{\eta _{s}}{T_{\perp s}}$,
where the pseudo-density $\eta _{s}$ is related to the leading-order species
number density $n_{s}$ to be determined in the next section.

It will be shown below that a basic consequence of this solution is that the
functions $\Lambda _{s}^{\left( 0\right) }\left( \Phi _{s}^{eff}\right) $
are related to the leading-order equilibrium fluid fields associated with $%
f_{bi-M,s}^{\prime }$. Notice that in Eq.(\ref{fbima}) no expansion has been
carried out in terms of the explicit dependences with respect to $\Phi
_{\ast s}$ and $m_{s}^{\prime }$. Recalling the expression of $m_{s}^{\prime
}$ given by its second-order approximation in Eq.(\ref{cipolla}), this means
that $f_{bi-M,s}^{\prime }$ carries an explicit contribution which is odd in
the parallel velocity $u^{\prime }$.

For completeness, we report also the expressions of the two functions $%
h_{s}^{\left( 1\right) }$ and $h_{s}^{\left( 2\right) }$:
\begin{eqnarray}
\sigma _{s}h_{s}^{\left( 1\right) } &\equiv &\sigma _{s}\frac{M_{s}}{2Z_{s}e}%
v^{2}\left( \frac{{Z_{s}e}\Phi _{s}^{eff}}{T_{\parallel s}}-\frac{1}{2}%
\right) \frac{\partial \ln T_{\parallel s}}{\partial \Phi _{s}^{eff}}  \notag
\\
&&+\sigma _{s}\frac{M_{s}}{2Z_{s}e}v^{2}\left[ \frac{\partial \ln \beta _{s}%
}{\partial \Phi _{s}^{eff}}-\mu _{0s}^{\prime }\frac{\partial \ln \alpha _{s}%
}{\partial \Phi _{s}^{eff}}\right] ,  \label{h1} \\
\varepsilon _{s}^{j}h_{s}^{\left( 2\right) } &\equiv &\varepsilon _{s}^{j}
\left[ \frac{\beta _{s}^{\left( 1\right) }}{\beta _{s}}-\frac{T_{\parallel
s}^{\left( 1\right) }}{2T_{\parallel s}}-\mu _{0s}^{\prime }\alpha
_{s}^{\left( 1\right) }\right]  \notag \\
&&+\varepsilon _{s}^{j}\frac{{Z_{s}e}\Phi _{s}^{eff}}{T_{\parallel s}^{2}}%
T_{\parallel s}^{\left( 1\right) },  \label{h2}
\end{eqnarray}%
where the expansion (\ref{cipolla}) for $m_{s}^{\prime }$ and the
leading-order approximation for $\Phi _{\ast s}$ have been used in this
case. We notice that, by construction, $\sigma _{s}h_{s}^{\left( 1\right) }$
and $\varepsilon _{s}^{j}h_{s}^{\left( 2\right) }$ are both even in $%
u^{\prime }$ and $\mathbf{w}^{\prime }$. Consistency of Eq.(\ref{h1})
requires that $\sigma _{s}\frac{M_{s}}{2Z_{s}e}v^{2}\left( \frac{{Z_{s}e}%
\Phi _{s}^{eff}}{T_{\parallel s}}-\frac{1}{2}\right) \frac{\partial \ln
T_{\parallel s}}{\partial \Phi _{s}^{eff}}\lesssim O\left( \sigma
_{s}\right) $, which implies that $T_{\parallel s}$ must be of the form $%
T_{\parallel s}=T_{\parallel s}\left( \sigma _{s}^{l}\Phi _{s}^{eff}\right) $%
, with $l\geq 1$, i.e., slowly-dependent on $\Phi _{s}^{eff}$. Similarly,
Eq.(\ref{h2}) demands that $\varepsilon _{s}^{j}\frac{{Z_{s}e}\Phi _{s}^{eff}%
}{T_{\parallel s}^{2}}T_{\parallel s}^{\left( 1\right) }\lesssim O\left(
\varepsilon _{s}^{j}\right) $, namely $\frac{T_{\parallel s}^{\left(
1\right) }}{T_{\parallel s}}\lesssim O\left( \sigma _{s}\right) $.

To conclude this section we remark that the SEPE regime requires that $\frac{%
{Z_{s}e}\Phi _{s}^{eff}}{T_{\parallel s}}\sim \frac{1}{O\left( \sigma
_{s}\right) }$. If, on the contrary, one imposes the asymptotic ordering $%
\frac{{Z_{s}e}\Phi _{s}^{eff}}{T_{\parallel s}}\sim O\left( \sigma
_{s}^{0}\right) $, it follows that for thermal particles it must be $\frac{%
\frac{M_{s}v^{2}}{2}}{{Z_{s}e}\Phi _{s}^{eff}}\sim O\left( \sigma
_{s}^{0}\right) $ too. Therefore, in such a case the $\sigma _{s}$-expansion
of the structure functions is no longer permitted. Therefore, when this
happens, the corresponding equilibrium KDF $f_{\ast s}$ remains generally
always non-Gaussian in the whole phase-space. The exception occurs when the
species structure functions are of the form (\ref{espan-1}) with $\Lambda
_{\ast s}^{\left( 0\right) }$ identically constant in phase-space. In fact,
when these conditions are met, $f_{\ast s}$ still admits a Chapman-Enskog
representation of the form (\ref{chens}) with $h_{s}^{\left( 1\right) }=0$
and the leading-order KDF coincides with the bi-Maxwellian KDF given by Eq.(%
\ref{fbima}) with uniformly-constant structure functions $\left\{ \Lambda
_{\ast s}^{\left( 0\right) }\right\} =\left\{ \beta _{s},\alpha
_{s},T_{\parallel s}\right\} $.

\bigskip

\section{Fluid moments}

In this section we evaluate the relevant approximations for the species
number density, flow velocity and pressure tensor carried by the equilibrium
KDF $f_{\ast s}$. These fluid fields can be explicitly analytically
evaluated in the case in which $f_{\ast s}$ admits the Chapman-Enskog
representation given by Eq.(\ref{chens}). This calculation requires
representing the same KDF in terms of the Newtonian state $\left( \mathbf{r},%
\mathbf{v}\right) $, i.e. making use of the inverse guiding-center
transformation to be determined from Eqs.(\ref{1-bis}) and (\ref{2-bis}).
This is achieved by letting%
\begin{eqnarray}
\mathbf{r}^{\prime } &=&\mathbf{r}-\varepsilon \mathbf{r}_{1}-\varepsilon
^{2}\mathbf{r}_{2}, \\
\mathbf{v} &=&u\mathbf{b}+\mathbf{w}+\mathbf{V}_{s}=u^{\prime }\mathbf{b}%
^{\prime }+\mathbf{w}^{\prime }+\mathbf{V}_{s}^{\prime }+\varepsilon \mathbf{%
v}_{1}^{\prime },
\end{eqnarray}%
where $\varepsilon \mathbf{r}_{1}=-\mathbf{w}\times \mathbf{b}/\Omega _{cs}$
and $\mathbf{V}_{s}\mathbf{=}\frac{c}{B}\mathbf{E}^{eff}\times \mathbf{b}$
is the $s$-species drift velocity. Correct to first-order this implies%
\begin{eqnarray}
u^{\prime } &=&u-\varepsilon \mathbf{r}_{1}\cdot \nabla \mathbf{b}\cdot
\mathbf{w}-\varepsilon \mathbf{r}_{1}\cdot \nabla \mathbf{b}\cdot \mathbf{V}%
-\varepsilon \mathbf{v}_{1}\cdot \mathbf{b}, \\
\mathbf{w}^{\prime } &=&\mathbf{w}+\varepsilon \left( \mathbf{r}_{1}\cdot
\nabla \mathbf{b}\cdot \mathbf{w}+\mathbf{r}_{1}\cdot \nabla \mathbf{b}\cdot
\mathbf{V}\right) \mathbf{b}  \notag \\
&&+\varepsilon u\mathbf{r}_{1}\cdot \nabla \mathbf{b}+\varepsilon \mathbf{r}%
_{1}\cdot \nabla \mathbf{V}-\varepsilon \mathbf{v}_{1}\cdot \left( \mathbf{I}%
-\mathbf{bb}\right) .
\end{eqnarray}%
As a consequence, one finds%
\begin{eqnarray}
\mu _{0s}^{\prime } &=&\mu _{0s}\left[ 1-\varepsilon \mathbf{w}\cdot \frac{%
\mathbf{b}\times \nabla \ln B}{\Omega _{c}}\right]  \notag \\
&&+\varepsilon \frac{2\mu _{0s}}{w^{2}}\left[ \mathbf{r}_{1}\cdot \nabla
\left( u\mathbf{b}+\mathbf{V}\right) \cdot \mathbf{w}+\mathbf{v}_{1}\cdot
\mathbf{w}\right] ,
\end{eqnarray}%
where $\mu _{0s}=\frac{M_{s}w^{2}}{2B}$, so that its gyrophase-average
evaluated at the position $\mathbf{r}$ yields:%
\begin{eqnarray}
\left\langle \mu _{0s}^{\prime }\right\rangle _{\phi } &=&\frac{1}{2\pi }%
\int_{0}^{2\pi }d\phi \mu _{0s}^{\prime }=  \notag \\
&=&\mu _{0s}\left[ 1-\varepsilon \frac{2}{\Omega _{c}B}\left[ u\frac{4\pi }{c%
}J_{\parallel }+\mathbf{B}\cdot \nabla \times \mathbf{V}\right] \right]
\notag \\
&&+\varepsilon \mu _{0s}\frac{2}{w^{2}}\Omega _{c}^{2}\left\langle \mathbf{r}%
_{2}\cdot \mathbf{r}_{1}\right\rangle ,
\end{eqnarray}%
with%
\begin{eqnarray}
\left\langle \mathbf{r}_{2}\cdot \mathbf{r}_{1}\right\rangle &=&\frac{w^{2}}{%
\Omega _{cs}^{3}}\left[ \mathbf{V}\cdot \nabla \ln B-\left( \underline{%
\mathbf{I}}-\mathbf{bb}\right) \mathbf{:}\nabla \mathbf{V}\right]  \notag \\
&=&\frac{w^{2}}{\Omega _{cs}^{3}}\left[ \frac{4\pi }{B^{2}}J_{\parallel
}E_{\parallel }^{eff}-\nabla \cdot \mathbf{V}-\frac{4\pi }{B^{\prime 2}}%
\mathbf{J}\cdot \mathbf{E}^{eff}\right] .
\end{eqnarray}%
Finally the magnetic moment perturbation $\varepsilon \mu _{1s}^{\prime }$
becomes simply to the same accuracy%
\begin{eqnarray}
\varepsilon \mu _{1}^{\prime } &\cong &-\varepsilon \mu _{0s}\frac{1}{B}%
\frac{1}{\Omega _{cs}}\left[ u\frac{4\pi }{c}J_{\parallel }+\mathbf{B}\cdot
\nabla \times \mathbf{V}\right]  \notag \\
&&-\varepsilon \mu _{0s}\frac{2}{\Omega _{cs}}\left[ \nabla \cdot \mathbf{V}+%
\frac{4\pi }{B^{2}}\mathbf{J}\cdot \mathbf{E}^{eff}\right]  \notag \\
&&+\varepsilon \mu _{0}\frac{1}{\Omega _{cs}}\frac{8\pi }{B^{2}}J_{\parallel
}E_{\parallel }^{eff}.
\end{eqnarray}

The leading-order expressions of the fluid fields carried by the equilibrium
Maxwellian-species KDF can in principle be evaluated with arbitrary
accuracy. To leading-order (with respect to all the relevant expansion
parameter) the species number density is determined by the velocity moment%
\begin{equation}
n_{s}=\int_{W}d^{3}vf_{bi-M,s}^{\left( 0\right) },
\end{equation}%
where%
\begin{equation}
f_{bi-M,s}^{\left( 0\right) }=\frac{\eta _{s}\exp \left\{ -\frac{{Z_{s}e}%
\Phi _{s}^{eff}}{T_{\parallel s}}\right\} }{\left( 2\pi /M_{s}\right)
^{3/2}T_{\perp s}T_{\parallel s}^{1/2}}\exp \left\{ -\frac{M_{s}v^{2}}{%
2T_{\parallel s}}-\mu _{0s}\alpha _{s}\right\}
\end{equation}%
is a local bi-Maxwellian KDF defined in terms of $\mu _{0s}$. It follows that%
\begin{equation}
n_{s}=\eta _{s}\exp \left\{ -\frac{{Z_{s}e}\Phi _{s}^{eff}}{T_{\parallel s}}-%
\frac{M_{s}V_{s}^{2}}{2\Delta _{T_{s}}}\frac{T_{\perp s}}{T_{\parallel s}}%
\right\} .  \label{ndensity}
\end{equation}%
The additional moment to be computed explicitly in terms of the KDF which is
needed for the evaluation of the Ampere equation is the total electric
current density, which in turn depends on the flow velocity. Let us compute
separately the perpendicular and parallel flow velocities and the
corresponding current densities. Concerning the perpendicular component of
the species flow velocity $\mathbf{U}_{\perp s}$, for local
Maxwellian-species its leading-order contribution is determined by the
following velocity moment%
\begin{equation}
\mathbf{U}_{\perp s}=\frac{1}{n_{s}}\int_{W}d^{3}v\left( \mathbf{w}+\mathbf{V%
}_{s}\right) f_{bi-M,s}^{\left( 0\right) },
\end{equation}%
which yields%
\begin{equation}
\mathbf{U}_{\perp s}=\left( 1-\frac{T_{\perp s}}{T_{\parallel s}}\right)
\mathbf{V}_{s}.
\end{equation}%
As a consequence, one finds that the perpendicular flow velocity is linearly
proportional to the effective electric-field drift velocity and vanishes in
case of isotropic species temperatures. The perpendicular current $\mathbf{J}%
_{\perp }^{\left( M\right) }$ generated by the Maxwellian-species is
therefore%
\begin{equation}
\mathbf{J}_{\perp }^{\left( M\right) }=\sum_{s\in \text{Maxwellian}%
}Z_{s}en_{s}\frac{T_{\perp s}}{\Delta _{T_{s}}}\mathbf{V}_{s}.
\label{geiperp}
\end{equation}%
Concerning the parallel flow velocity, this is in principle determined for
Maxwellian-species by the whole Chapman-Enskog representation as%
\begin{equation}
U_{\parallel s}=\frac{1}{n_{s}}\int_{W}d^{3}vuf_{bi-M,s}^{\prime }\left[
1+\sigma _{s}h_{s}^{\left( 1\right) }+\varepsilon _{s}^{j}h_{s}^{\left(
2\right) }\right] .
\end{equation}%
Since $\mathbf{V}_{s}$ is purely perpendicular, while both $h_{s}^{\left(
1\right) }$ and $h_{s}^{\left( 2\right) }$ are even in $u$, it follows that
the leading-order contribution is necessarily provided by%
\begin{equation}
U_{\parallel s}=-\varepsilon _{s}\frac{1}{n_{s}}\int_{W}d^{3}vuf_{bi-M,s}^{%
\left( 0\right) }\left( \alpha _{s}\Delta \mu _{s}^{1}\right) ,
\end{equation}%
with $\Delta \mu _{s}^{1}=-u\mu _{0s}\frac{3}{\Omega _{c}B}\frac{4\pi }{c}%
J_{\parallel }$. Explicit calculation for the local Maxwellian-species gives%
\begin{equation}
U_{\parallel s}=\varepsilon _{s}\frac{1}{\Omega _{c}B}\frac{6\pi }{c}%
J_{\parallel }\left[ 2\frac{T_{\parallel s}}{M_{s}}+V^{2}\frac{T_{\perp s}}{%
T_{\parallel s}}\right] \frac{T_{\perp s}}{\Delta _{T_{s}}},
\end{equation}%
which vanishes identically only if $J_{\parallel }\frac{1}{\Delta _{T_{s}}}%
=0 $, namely if either the total parallel current density is null or the
species temperature is isotropic. The total parallel current density $%
J_{\parallel }^{\left( M\right) }$ carried by all the Maxwellian species is
therefore $J_{\parallel }^{\left( M\right) }=\gamma J_{\parallel }$, with%
\begin{equation}
\gamma \equiv \frac{6\pi }{B^{2}}\sum_{s\in \text{Maxwellian}}\varepsilon
_{s}n_{s}\left[ 2T_{\parallel s}+M_{s}V^{2}\frac{T_{\perp s}}{T_{\parallel s}%
}\right] \frac{T_{\perp s}}{\Delta _{T_{s}}}.
\end{equation}%
Hence $J_{\parallel }^{\left( M\right) }$ can be generally non-zero even for
quasi-neutral plasmas only provided there is an additional source for $%
J_{\parallel }$. At equilibrium, there is only one possible physical
mechanism consistent with the solution indicated above for $f_{\ast s}$.
This occurs when there exists locally non-Maxwellian populations of
particles described by KDFs of the form (\ref{sol2}), which must co-exist in
the same domain of the plasma. Denoting by $J_{\parallel }^{\left(
non-M\right) }$ the corresponding current density it follows that the total
current $J_{\parallel }$ is necessarily given by%
\begin{equation}
J_{\parallel }=\frac{J_{\parallel }^{\left( non-M\right) }}{1-\gamma },
\label{paralcur}
\end{equation}%
so that $J_{\parallel }$ can be strongly amplified when $\gamma \sim 1$,
i.e., for high-beta plasmas characterized by strong temperature anisotropies
of the Maxwellian species.

Finally, the leading-order pressure tensor for local Maxwellian-species is
found to be given by the velocity moment%
\begin{equation}
\underline{\underline{\Pi }}_{s}=M_{s}\int_{W}d^{3}v\left( \mathbf{v}-%
\mathbf{U}_{\perp s}\right) \left( \mathbf{v}-\mathbf{U}_{\perp s}\right)
f_{bi-M,s}^{\left( 0\right) },
\end{equation}%
which yields%
\begin{equation}
\underline{\underline{\Pi }}_{s}=p_{\perp s}\underline{\mathbf{I}}+\left(
p_{\parallel s}-p_{\perp s}\right) \mathbf{bb},  \label{tensor-pressure anis}
\end{equation}%
where $p_{\perp s}\equiv n_{s}T_{\perp s}$ and $p_{\parallel s}\equiv
n_{s}T_{\parallel s}$ represent the leading-order species perpendicular and
parallel pressures for Maxwellian-species.

\bigskip

\section{The Poisson equation for ES potential}

In this section we analyze the Poisson equation which determines the ES
potential $\Phi $ in terms of the number density $n_{s}$ determined in Eq.(%
\ref{ndensity}) for Maxwellian-species. This analysis provides also the
constraint conditions for the realizability of Maxwellian-like kinetic
equilibria and the existence of the related Chapman-Enskog expansion of the
type given by Eq.(\ref{chens}).

We first consider the case in which quasi-neutrality condition applies.
Then, for local-Maxwellian species the Poisson equation requires to
leading-order%
\begin{equation}
\sum_{s\in \text{Maxwellian}}Z_{s}en_{s}=0.
\end{equation}%
For a two-species ion-electron plasma ($s=i,e$), substituting for $n_{s}$
from Eq.(\ref{ndensity}) gives%
\begin{eqnarray}
&&\left. \eta _{i}\left\vert Z_{i}e\right\vert \exp \left\{ -\frac{{Z_{i}e}%
\Phi _{i}^{eff}}{T_{\parallel i}}-\frac{M_{i}V_{i}^{2}}{2\Delta _{T_{i}}}%
\frac{T_{\perp i}}{T_{\parallel i}}\right\} \right.  \notag \\
&&\left. =\eta _{e}\left\vert e\right\vert \exp \left\{ -\frac{{e}\Phi
_{e}^{eff}}{T_{\parallel e}}-\frac{M_{e}V_{e}^{2}}{2\Delta _{T_{e}}}\frac{%
T_{\perp e}}{T_{\parallel e}}\right\} .\right.
\end{eqnarray}%
Recalling that $\Phi _{s}^{eff}\equiv \Phi +\frac{M_{s}}{Z_{s}e}\Phi _{G}$
one obtains to leading-order%
\begin{equation}
\Phi =\frac{X_{ie}-\left[ \frac{M_{i}}{T_{\parallel i}}-\frac{M_{e}}{%
T_{\parallel e}}\right] \Phi _{G}}{\left[ \frac{\left\vert {Z_{i}e}%
\right\vert }{T_{\parallel i}}+\frac{\left\vert {e}\right\vert }{%
T_{\parallel e}}\right] },  \label{ES}
\end{equation}%
where%
\begin{equation}
X_{ie}\equiv \ln \left[ \frac{\eta _{i}\left\vert Z_{i}e\right\vert }{\eta
_{e}\left\vert e\right\vert }\right] -\frac{M_{i}V_{i}^{2}}{2\Delta _{T_{i}}}%
\frac{T_{\perp i}}{T_{\parallel i}}+\frac{M_{e}V_{e}^{2}}{2\Delta _{T_{e}}}%
\frac{T_{\perp e}}{T_{\parallel e}}.
\end{equation}%
Eq.(\ref{ES}) represents an implicit solution for the ES potential $\Phi $.
In fact both the species temperatures and pseudo-densities on the rhs of the
same equation can still depend functionally on $\Phi _{s}^{eff}$, according
to the kinetic constraint prescribed on the equilibrium solution. In
addition, the quantity $X_{ie}$ depends also on the drift velocity $%
V_{s}^{2} $, which is again still a function of the same effective potential.

We first notice that for astrophysical gravitationally-bound plasmas in
which the ordering condition $\left\vert \frac{M_{i}\Phi _{G}}{T_{\parallel
i}}\right\vert \sim \frac{1}{O\left( \sigma _{i}\right) }$ holds, it then
follows that also $\left\vert \frac{{Z_{s}e}\Phi }{T_{\parallel s}}%
\right\vert \sim \frac{1}{O\left( \sigma _{s}\right) }$ applies. This
conclusion is consistent with the requirement of SEPE regime, so that in
this case the $\sigma _{s}$-ordering expansion is valid and the equilibrium
KDF admits the Chapman-Enskog representation (\ref{chens}) with
non-vanishing $h_{s}^{\left( 1\right) }$. On the other hand, for
astrophysical plasmas in which $\left\vert \frac{M_{i}\Phi _{G}}{%
T_{\parallel i}}\right\vert \sim O\left( \sigma _{i}^{0}\right) $ and for
laboratory plasmas for which $\Phi _{G}$ is negligible, since $O\left(
X_{ie}\right) \lesssim O\left( \sigma _{i}^{0}\right) $, then Eq.(\ref{ES})
implies that $\Phi $ satisfies necessarily the ordering $\left\vert \frac{{%
Z_{s}e}\Phi }{T_{\parallel s}}\right\vert \lesssim O\left( \sigma
_{s}^{0}\right) $. This condition violates the requirement for the existence
of the SEPE regime, so that in this case the $\sigma _{s}$-expansion does
not apply. As discussed in Section 7, this conclusion implies in turn that
the equilibrium KDF $f_{\ast s}$ remains generally non-Gaussian in the whole
phase-space, unless the corresponding structure functions are taken to be
identically constant in the whole phase-space. In this case $f_{\ast s}$ can
still be expressed to leading-order in terms of the bi-Maxwellian KDF given
by Eq.(\ref{fbima}), but with uniformly-constant structure functions, while
the energy-correction contributions $h_{s}^{\left( 1\right) }$ in the
Chapman-Enskog representation (\ref{chens}) vanish identically. Of course
this does not generally imply a condition of constant fluid fields, provided
the magnetic and electric fields are spatially non-uniform.

Let us now address the case of non-neutral plasmas for local-Maxwellian
species. For simplicity, let us consider the case in which there exists a
single-species plasma. The determination of the ES potential now necessarily
requires integration of the Poisson equation. Invoking again Eq.(\ref%
{ndensity}) for the species number density, one obtains the non-linear
differential equation%
\begin{equation}
\nabla ^{2}\Phi =-4\pi Z_{s}e\eta _{s}\exp \left\{ -\frac{{Z_{s}e}\Phi
_{s}^{eff}}{T_{\parallel s}}-\frac{M_{s}V_{s}^{2}}{2\Delta _{T_{s}}}\frac{%
T_{\perp s}}{T_{\parallel s}}\right\} .  \label{nonnneu}
\end{equation}%
In the presence of an external gravitational field, let us assume that $\Phi
_{G}$ satisfies the ordering $\left\vert \frac{M{_{s}}\Phi _{G}}{%
T_{\parallel s}}\right\vert \sim \frac{1}{O\left( \sigma _{s}\right) }$. In
this case one can infer that $\Phi $ must remain always exponentially small,
both for ${Z_{s}e>0}$ or ${Z_{s}e<0}$. When this happens, a local Maxwellian
equilibrium solution of the type obtained above exists, consistent with the
ordering assumptions introduced for the SEPE regime. Instead, if $\Phi _{G}$
is negligible (as in the case of laboratory plasmas) and we impose at the
same time the orderings $\left\vert \frac{{Z_{s}e}\Phi }{T_{\parallel s}}%
\right\vert \sim \frac{1}{O\left( \sigma _{s}\right) }$ and $\left\vert
\frac{Z_{s}e\eta _{s}}{\nabla ^{2}\Phi }\right\vert \sim O\left( \sigma
_{s}^{k}\right) $, with $k\geq 1$, it follows that a local Maxwellian
equilibrium with non-uniform structure functions cannot generally exist. In
fact in this case Eq.(\ref{nonnneu}) would demand for a generic species $s$
that $O\left( \sigma _{s}^{k}\right) \sim e^{-\frac{1}{O\left( \sigma
_{s}\right) }}$, and therefore that $k\rightarrow 0$. As a consequence also
in the case of non-neutral plasmas the only admissible non-symmetric
Maxwellian-like kinetic equilibrium for laboratory plasmas requires constant
structure functions (no $\sigma _{s}$-expansion).

\bigskip

\section{Ampere's equation and the kinetic dynamo}

In this section we investigate the solubility conditions posed by the Ampere
equation. Let us assume that the magnetic field admits locally a
representation in terms of Clebsch potentials $\left( \psi ,Q\right) $ of
the form $\mathbf{B}=\nabla \psi \times \nabla Q$.

For definiteness, let us consider first the case of flow-through equilibria
in which the magnetic surfaces $\psi \left( \mathbf{r}\right) =const.$ and $%
Q\left( \mathbf{r}\right) =const.$ are open (i.e., unbounded) in the
Euclidean space $%
\mathbb{R}
^{3}$ occupied by the plasma. Then the only constraint condition to be taken
into account is given by the requirement of solenoidal current density%
\begin{equation}
\nabla \cdot \mathbf{J}=0.  \label{solenoid}
\end{equation}%
For a stationary plasma, the constraint (\ref{solenoid}) is necessarily
already satisfied by the exact equilibrium KDF $f_{\ast s}$ \cite%
{Cr2011,Cr2011a}. More precisely, the latter is fulfilled either identically
or asymptotically depending whether $f_{\ast s}$ is a first integral or an
adiabatic invariant.

As a basic consequence, when the Chapman-Enskog expansion (\ref{chens}) is
adopted for $f_{\ast s}$ and consequently $\mathbf{J}\equiv \sum_{s}\mathbf{J%
}_{s}$ is expanded in power series of $\varepsilon _{s}$ and $\sigma _{s}$,
the previous equation generally implies a relationship between the various
series perturbative contributions. For example, let us assume that for each
species $\mathbf{J}_{s}$ is determined asymptotically up to first-order, so
that $\mathbf{J}_{s}\cong \mathbf{J}_{s}^{\left( 0\right) }+\varepsilon _{s}%
\mathbf{J}_{\varepsilon s}^{\left( 1\right) }+\sigma _{s}\mathbf{J}_{\sigma
s}^{\left( 1\right) }$. Then one obtains\ that%
\begin{equation}
\nabla \cdot \sum_{s}\mathbf{J}_{s}^{\left( 0\right) }\cong -\nabla \cdot
\sum_{s}\left( \varepsilon _{s}\mathbf{J}_{\varepsilon s}^{\left( 1\right)
}+\sigma _{s}\mathbf{J}_{\sigma s}^{\left( 1\right) }\right) .  \label{amp}
\end{equation}
Notice that, as a consequence, no physical constraints can possibly arise on
the functional form of $\mathbf{J}_{s}^{\left( 0\right) }$ or the
perturbations $\mathbf{J}_{\varepsilon s}^{\left( 1\right) }$ and $\mathbf{J}%
_{\sigma s}^{\left( 1\right) }$. This shows that, in order to satisfy the
constraint condition (\ref{amp}) in asymptotic sense, the corresponding
Ampere's equation must also include first-order currents. Therefore in such
a case one obtains the asymptotic equation%
\begin{equation}
\nabla \times \mathbf{B}=\frac{4\pi }{c}\sum_{s}\left[ \mathbf{J}%
_{s}^{\left( 0\right) }+\varepsilon _{s}\mathbf{J}_{\varepsilon s}^{\left(
1\right) }+\sigma _{s}\mathbf{J}_{\sigma s}^{\left( 1\right) }\right] .
\end{equation}%
Basic implications are:

1) The possible existence of a kinetic dynamo effect also in the case of
spatially non-symmetric EM fields, which may be responsible for the
self-generation of equilibrium magnetic field by the plasma itself. This
conclusion is in agreement with the predictions obtained in the case of
symmetric systems (see Refs.\cite{Cr2010,Cr2011,Cr2011a}).

2)\ The possible occurrence of a \textquotedblleft twist\textquotedblright\
effect which characterizes the equilibrium magnetic field. This is produced
by the combined effect of perpendicular and parallel plasma currents in
spatially non-symmetric equilibria.

3) For gravitationally-bound astrophysical plasma for which the ordering $%
\left\vert \frac{M{_{s}}\Phi _{G}}{T_{\parallel s}}\right\vert \sim \frac{1}{%
O\left( \sigma _{s}\right) }$ applies, the leading-order current-density
carried by local-Maxwellian species is given by Eq.(\ref{geiperp}), while
the parallel current density is necessarily of first-order and is given in
terms of Eq.(\ref{paralcur}).

4) The treatment of laboratory plasmas is different, because the drift
velocity $\mathbf{V}_{s}$ and the corresponding current density are always
of $O\left( \varepsilon _{s}\right) $ with respect to $v_{ths}$, while at
the same time the structure functions must be constant.

5) In both cases 3) and 4), the parallel current density can only arise in
the presence of locally non-Maxwellian species.

Let us now consider the solubility conditions which arise in the case in
which the magnetic surfaces $\psi \left( \mathbf{r}\right) =const.$ are
locally closed and nested. We stress that, in principle, also this case can
be regarded as a flow-through equilibrium provided suitable plasma sources
are set up in order to compensate for the possible occurrence of
corresponding loss-cones. For definiteness, to this aim we introduce again
the magnetic coordinates $\left( \psi ,\varphi ,\vartheta \right) $ and
identify $Q=\varphi -q\left( \psi \right) \vartheta $, with $q\left( \psi
\right) $ being the rotational transform. For definiteness, we limit our
analysis to the periodicity conditions following from the constraint
equation (\ref{solenoid}) indicated above. Then, on a rational $\psi $%
-magnetic surface in which $q\left( \psi \right) =\frac{m}{n}$, with $m\in
\mathbb{N}
$ and $n\in
\mathbb{N}
_{0}$, Eq.(\ref{solenoid}) implies%
\begin{equation}
\int_{0}^{2\pi m}d\vartheta J_{p}\nabla \cdot \mathbf{J}_{\perp }=0,
\end{equation}%
where $J_{p}$ is the Jacobian $J_{p}=\frac{1}{\left\vert \mathbf{B}\cdot
\nabla \vartheta \right\vert }$. Introducing for $\mathbf{J}_{\perp }$ the
perturbative expansion indicated above, it follows%
\begin{eqnarray}
&&\left. \int_{0}^{2\pi m}d\vartheta J_{p}\nabla \cdot \sum_{s}\mathbf{J}%
_{\perp s}^{\left( 0\right) }=\right.  \notag \\
&&\left. =-\int_{0}^{2\pi m}d\vartheta J_{p}\nabla \cdot \sum_{s}\left(
\varepsilon _{s}\mathbf{J}_{\perp \varepsilon s}^{\left( 1\right) }+\sigma
_{s}\mathbf{J}_{\perp \sigma s}^{\left( 1\right) }\right) .\right.
\label{grad}
\end{eqnarray}%
This equation is qualitative different from the customary solubility
condition usually considered in the literature (see for example Refs.\cite%
{Grad58,kuls,kuls2,Newcomb}) for the following reasons:

1)\ If we restrict ourselves to the case in which the perpendicular current
density is only generated by local-Maxwellian species, so that $\mathbf{J}%
_{\perp }=\mathbf{J}_{\perp }^{\left( M\right) }$, then both the
leading-order current density $\sum_{s}\mathbf{J}_{\perp s}^{\left( 0\right)
}$ as well as the first-order perturbations $\sum_{s}\left( \varepsilon _{s}%
\mathbf{J}_{\perp \varepsilon s}^{\left( 1\right) }+\sigma _{s}\mathbf{J}%
_{\perp \sigma s}^{\left( 1\right) }\right) $ arise only in the presence of
temperature anisotropy (i.e., they all vanish identically in the case of
isotropic species temperature).

2) The equation is asymptotic and therefore the lhs vanishes only when
first-order corrections in $\varepsilon _{s}$ and $\sigma _{s}$ are
neglected.

For astrophysical plasmas the solubility condition (\ref{grad}) can always
be satisfied by suitable prescription of the spatially non-uniform structure
functions, for example the ion perpendicular temperature $T_{\perp i}$. In
fact, the latter for an arbitrary species is generally of the form $T_{\perp
s}=T_{\perp s}\left( B,\Phi _{s}^{eff}\right) $. If $B$ and $\Phi _{s}^{eff}$
are considered independent it follows that $T_{\perp i}=\widehat{T}_{\perp
i}\left( \mathbf{r}\right) $ remains arbitrary, so that it can always be
determined by Eq.(\ref{grad}). In the case of laboratory plasmas instead $%
T_{\perp s}=T_{\perp s}\left( B\right) $, which means that it cannot longer
be used as a free parameter. On the other hand in this case the
perpendicular current contains only first-order terms, which can include
also contributions of locally non-Maxwellian species or perturbations of
local Maxwellian species associated to the terms $\varepsilon
_{s}^{j}\Lambda _{s}^{\left( 1\right) }\left( \Phi _{s}^{eff},\mu
_{0s}^{\prime }\right) $ in Eq.(\ref{taylor lambda}). In both cases, in view
of the general form of the functional dependences of the structure
functions, the latter can always be prescribed in such a way to fulfill Eq.(%
\ref{grad}).

\bigskip

\section{Concluding remarks}

In this paper the construction of spatially non-symmetric kinetic equilibria
has been addressed. The theory has been developed in the framework of the
Vlasov-Maxwell treatment and applies to non-relativistic collisionless
magnetized plasmas subject to both electromagnetic and gravitational fields.

The result is based on a non-perturbative formulation of guiding-center
theory for single-particle dynamics, carried out in terms of a variational
Lagrangian formulation obtained by introducing a finite phase-space
gyrokinetic transformation of the particle Newtonian state. The theory holds
for non-relativistic single charged particles moving in external
non-homogeneous suitably-smooth electromagnetic and gravitational fields.
The conditions of existence of the gyrokinetic transformation have been
determined in terms of solubility conditions for the corresponding
constraint equations which are determined by the Lagrangian formulation. As
a consequence of the symmetry properties of the Hamilton variational
principle, an exact formal representation of the magnetic moment has been
obtained, which recovers the form of a particle canonical momentum and is
proved to be, under suitable assumptions, a local phase-space first-integral
of motion. As an application, a second-order Larmor-radius expansion for
gyrokinetic theory has been presented. The corresponding approximation for
the magnetic moment has been explicitly calculated \textquotedblleft a
posteriori\textquotedblright\ and proved to represent an adiabatic invariant
of prescribed order.

Then, exact quasi-stationary solutions of the Vlasov equation for spatially
non-symmetric configurations have been determined. These are expressed as
functions of the relevant particle adiabatic invariants, namely the total
particle energy and the magnetic moment. An explicit representation in terms
of generalized bi-Maxwellian distributions has been given, proving that the
solution is generally non-Gaussian. In terms of a suitable perturbative
theory holding for local-Maxwellian species, the equilibria thus obtained
have been shown to admit a Chapman-Enskog representation. Analytical
calculations of the corresponding fluid moments has been obtained, which
include the species number density, flow velocity and pressure tensor. As a
notable outcome, it has been pointed out that non-symmetric kinetic
equilibria of this kind admit both perpendicular and parallel components of
the flow velocity. For Maxwellian species, both exist provided the
temperature is non-sotropic, while the parallel component requires in
addition a non-vanishing equilibrium parallel current density. The
constraints posed by the Maxwell equations have been analyzed and proved to
be satisfied both in the case of astrophysical and laboratory plasmas.
First, from the Poisson equation it has been shown that these solutions are
consistent with quasi-neutrality condition as well as with configurations of
non-neutral plasmas. Then, analysis of the Ampere equation has revealed the
existence of kinetic dynamo effects, namley the possibility of
self-generation of quasi-stationary magnetic fields by the same equilibrium
plasma currents.

The theory presented here provides a useful background for possible
theoretical investigations of a wide range of plasma kinetic equilibria and
corresponding stability analyses. In particular it permits applications for
the study of spatially non-symmetric plasmas in both laboratory contexts
\cite{Co09,Co10} and astrophysical scenarios, including non-symmetric
structures in accretion disc plasmas around compact objects \cite%
{spot1,spot2} and in the solar corona, like solar arcades \cite{arc1},
prominences \cite{pro1} and loops \cite{loop1}.

\begin{acknowledgments}
This work has been partly developed in the framework of MIUR (Italian
Ministry for Universities and Research) PRIN Research Program
\textquotedblleft Problemi Matematici delle Teorie Cinetiche e
Applicazioni\textquotedblright , University of Trieste, Italy.
\end{acknowledgments}

\section{Appendix: Hamilton variational principle and GK Euler-Lagrange
equations}

The Hamilton variational principle in superabundant hybrid (i.e.,
non-canonical) variables has first been given in Ref.\cite{Littke1983}. It
amounts to introduce the Hamilton variational functional of the form%
\begin{eqnarray}
J\left( \mathbf{r},\mathbf{v}\right) &=&\int_{t_{1}}^{t_{2}}\Lambda \left(
\mathbf{r},\frac{d}{dt}\mathbf{r},\mathbf{v},t\right) =  \notag \\
&=&\int_{t_{1}}^{t_{2}}dt\mathcal{L}\left( \mathbf{r}\left( t\right) ,\frac{d%
}{dt}\mathbf{r}\left( t\right) ,\mathbf{v}\left( t\right) ,t\right) ,
\end{eqnarray}%
with $\mathcal{L}\left( \mathbf{r},\frac{d}{dt}\mathbf{r},\mathbf{v}%
,t\right) \equiv \frac{d}{dt}\mathbf{r}\cdot \left[ M_{s}\mathbf{v}+\frac{%
Z_{s}e}{c}\mathbf{A}\right] -\mathcal{H}_{s}$ denoting the Lagrangian
expressed in terms of the hybrid variables $\mathbf{r}$ and $\mathbf{v}$,
which are considered as independent and $\mathcal{H}_{s}$ the
single-particle energy given by Eq.(\ref{energii}). The extremal curves $%
\mathbf{r}\left( t\right) $ and $\mathbf{v}\left( t\right) $, solutions of
the Hamilton variational principle%
\begin{equation}
\delta J\left( \mathbf{r},\mathbf{v}\right) =0,  \label{hvp}
\end{equation}%
which have null variations at the extrema $t_{1}$, $t_{2}$, satisfy the
Euler-Lagrange equations%
\begin{eqnarray}
&&\left. \frac{\delta J\left( \mathbf{r},\mathbf{v}\right) }{\delta \mathbf{v%
}}=\frac{\partial \mathcal{L}}{\partial \mathbf{v}}=M_{s}\left[ \frac{d}{dt}%
\mathbf{r}-\mathbf{v}\right] =\mathbf{0},\right. \\
&&\left. \frac{\delta J\left( \mathbf{r},\mathbf{v}\right) }{\delta \mathbf{r%
}}=\frac{\partial \mathcal{L}}{\partial \mathbf{r}}-\frac{d}{dt}\frac{%
\partial \mathcal{L}}{\partial \frac{d}{dt}\mathbf{r}}=\right.  \notag \\
&&\left. =Z_{s}e\left( \mathbf{E}^{eff}+\frac{\mathbf{v}}{c}\times \mathbf{B}%
\right) -M_{s}\frac{d}{dt}\mathbf{v=0}.\right.
\end{eqnarray}%
The same variational principle holds also in arbitrary hybrid variables.
Therefore, introducing in particular a phase-space diffeomorfism of the form
(\ref{gktr}), among the transformed variables one can always set $%
y_{4}^{\prime }=\mathcal{H}_{s}$ and $y_{3}^{\prime }=p_{\phi ^{\prime }s}$,
where $p_{\phi ^{\prime }s}$ is the canonical momentum conjugate to the
generalized coordinate $\phi ^{\prime }$ introduced in Eq.(\ref{gktr}).

The corresponding Euler-Lagrange equations expressed in terms of the GK
transformed state $\mathbf{z}^{\prime }$ defined in Section 3 (see Eq.(\ref%
{gktr})) are determined by the GK differential 1-form $\Lambda _{1}\left(
\mathbf{z}^{\prime },\frac{d}{dt}\mathbf{z}^{\prime }\right) $ given by Eq.(%
\ref{lambda1}). The Hamilton variational principle (\ref{hvp}) delivers the
Euler-Lagrange equations%
\begin{equation}
\frac{\delta J\left( \mathbf{z}^{\prime }\right) }{\delta \mathbf{z}^{\prime
}}=\frac{\partial \mathcal{L}_{1}}{\partial \mathbf{z}^{\prime }}-\frac{d}{dt%
}\frac{\partial \mathcal{L}_{1}}{\partial \frac{d}{dt}\mathbf{z}^{\prime }}%
=0.
\end{equation}%
For $\mathbf{z}^{\prime }\equiv \left( \mathbf{r}^{\prime },\mathcal{H}%
_{s},p_{\phi ^{\prime }s},\phi ^{\prime }\right) $ one obtains respectively:%
\begin{equation}
\frac{\delta J\left( \mathbf{z}^{\prime }\right) }{\delta \mathcal{H}_{s}}=%
\frac{\partial \mathcal{L}_{1}}{\partial \mathcal{H}_{s}}=\frac{d\mathbf{r}%
^{\prime }}{dt}\cdot \frac{\partial \left\langle \mathbf{\Gamma }_{\mathbf{r}%
^{\prime }}-\nabla ^{\prime }R\right\rangle }{\partial \mathcal{H}_{s}}-1=0,
\end{equation}%
\begin{equation}
\frac{\delta J\left( \mathbf{z}^{\prime }\right) }{\delta p_{\phi ^{\prime
}s}}=\frac{\partial \mathcal{L}_{1}}{\partial p_{\phi ^{\prime }s}}=\frac{d%
\mathbf{r}^{\prime }}{dt}\cdot \frac{\partial \left\langle \mathbf{\Gamma }_{%
\mathbf{r}^{\prime }}-\nabla ^{\prime }R\right\rangle }{\partial p_{\phi
^{\prime }s}}+\frac{d\phi ^{\prime }}{dt}=0,
\end{equation}%
\begin{equation}
\frac{\delta J\left( \mathbf{z}^{\prime }\right) }{\delta \phi ^{\prime }}=-%
\frac{d}{dt}\frac{\partial \mathcal{L}_{1}}{\partial \frac{d}{dt}\phi
^{\prime }}=-\frac{dp_{\phi ^{\prime }s}}{dt}=0,
\end{equation}%
\begin{equation}
\frac{\delta J\left( \mathbf{z}^{\prime }\right) }{\delta \mathbf{r}^{\prime
}}=\frac{\partial \mathcal{L}_{1}}{\partial \mathbf{r}^{\prime }}-\frac{d}{dt%
}\frac{\partial \mathcal{L}_{1}}{\partial \frac{d}{dt}\mathbf{r}^{\prime }}=%
\frac{d\mathbf{r}^{\prime }}{dt}\times \mathbf{B}^{\ast }=\mathbf{0},
\end{equation}%
where $\mathbf{B}^{\ast }\equiv \frac{c}{Z_{s}e}\left[ \nabla ^{\prime
}\times \left\langle \mathbf{\Gamma }_{\mathbf{r}^{\prime }}-\nabla ^{\prime
}R\right\rangle \right] $ is the effective magnetic field and $\left\langle
\mathbf{\Gamma }_{\mathbf{r}^{\prime }}-\nabla ^{\prime }R\right\rangle $
its corresponding effective vector potential. It is immediate to show that,
under the assumption of constant particle energy ($\mathcal{H}_{s}=const.$),
the previous equations recover the correct asymptotic approximations when a
LRE is performed (see Section 2).

\end{document}